\documentstyle[12pt]{article}
\input{psfig}
\begin{document}
\newcommand{\pom}{I\!\! P}
\newcommand{\nnn}{\noindent}
\title{ \huge \bf
Factorization and Scaling\\in Hadronic Diffraction}
\author{\large
K. Goulianos \\
(dino@physics.rockefeller.edu) \\
and \\
J. Montanha$^\dagger$ \\
(montanha@d0rio0.fnal.gov) \\ \\
The Rockefeller University \\
1230 York Avenue \ New York \ NY \ 10021 \ USA}
\date{May 21, 1998\\
Rockefeller University Preprint:  RU 97/E-43}
\maketitle

\begin{abstract}
In standard Regge theory, the contribution of the tripe-pomeron amplitude to 
the $t=0$ differential cross section for
single diffraction dissociation has the form
$d^2\sigma/dM^2 dt|_{t=0}\sim (s/s_0)^{2[\alpha_{\pom}(0)-1]}
/(M^2)^{\alpha_{\pom}(0)}$, where $\alpha_{\pom}(t)$ is the pomeron trajectory.
For $\alpha_{\pom}(0)>1$, this form,  
which is based on factorization, does not scale with energy.  From an
analysis of $pp$ and $\bar{p} p$ data from fixed target to collider
energies, we find that such scaling actually holds, signaling a breakdown 
of factorization.
Phenomenologically, this result can be obtained 
from a scaling law in diffraction, 
which is embedded in the hypothesis of pomeron flux renormalization 
introduced to unitarize the triple pomeron amplitude.  
\end{abstract}


$\dagger$ Fellow of CNPq, Brazil.

\newpage

\section{Introduction}
\nnn In Regge theory, the high energy behavior of hadronic cross sections is 
dominated by pomeron exchange~\cite{KG,DLT}. 
For a simple pomeron pole, the $pp$ elastic, 
total, and single diffractive (SD) cross sections can be written as
\begin{equation}
\frac{d\sigma_{el}}{dt}=\frac{\beta^4_{\pom pp}(t)}{16\pi}\;
{\left(\frac{s}{s_0}\right)}^{2[\alpha_{\pom}(t)-1]}
,
\label{elastic}
\end{equation}
\begin{equation}
\sigma_T(s)=\beta^2_{\pom pp}(0)
\left(\frac{s}{s_0}\right)^{\alpha_{\pom}(0)-1}
,
\label{total}
\end{equation}
\begin{equation}
\frac{d^2\sigma_{sd}}{d\xi dt}=
\frac{{\beta_{\pom pp}^2(t)}}{16\pi}\;\xi^{1-2\alpha_{\pom}(t)}
\left[\beta_{\pom pp}(0)\,g(t)
\;\left(\frac{s'}{s_0}\right)^{\alpha_{\pom}(0)-1}\right],
\label{diffractive}
\end{equation}

\nnn where $\alpha_{\pom}(t)=\alpha_{\pom}(0)+\alpha' t=(1+\epsilon)+\alpha' t$
is the pomeron Regge 
trajectory, $\beta_{\pom pp}(t)$ is the coupling of the pomeron to the proton,
$g(t)$ is the triple-pomeron coupling, $s'=M^{2}$ is the $\pom-p$ center of
mass energy squared, $\xi = 1-x_{F}=s'/s=M^2/s$ is the fraction of
the momentum of the proton carried by the pomeron, and $s_0$ is an energy
scale parameter, which is assumed throughout this paper to be 1~GeV$^2$ 
unless appearing explicitely.

In analogy with Eq.~\ref{total}, the term in brackets in (\ref{diffractive})
is identified as the $\pom-p$ total cross section, 
\begin{equation}
\sigma_T^{\pom p}(s',t)=\sigma_T^{\pom p}(s')=
\beta_{\pom pp}(0)\,g(0)
\,\left(\frac{s'}{s_0}\right)^{\alpha_{\pom}(0)-1}
\equiv \sigma_0^{\pom p}\,\left(\frac{s'}{s_0}\right)^{\alpha_{\pom}(0)-1}\,,
\label{pom-total}
\end{equation}
where we have used $g(t)=g(0)$, since it was found experimentally that 
$g(t)$ is independent of $t$~\cite{KG}.
The remaining factor in (\ref{diffractive}), namely
\begin{equation}
f_{\pom /p}(\xi,t)\equiv \frac{{\beta_{\pom pp}^2(t)}}{16\pi}\;
\xi^{1-2\alpha_{\pom}(t)}
\equiv K\,\xi^{1-2\alpha_{\pom}(t)}\,F^2(t),
\label{flux}
\end{equation}
where $K\equiv \beta_{\pom pp}^2(0)/16\pi$, 
is interpreted as the ``pomeron flux". 
Thus, $pp$ diffraction dissociation can be viewed as a process in which 
pomerons emitted by one of the protons interact with the other proton~\cite{IS}.

The function $F(t)$ represents the 
proton form factor, which is obtained from elastic scattering. 
At small $t$, $F^2(t)\approx e^{4.6t}$~\cite{R}. However, this simple 
exponential expression underestimates the cross section at large $t$. 
Donnachie and Landshoff proposed~\cite{DLF} 
that the appropriate 
form  factor for $pp$ elastic and diffractive scattering 
is the isoscalar form factor measured in electron-nucleon 
scattering, namely 
\begin{equation}
F_1(t)=\frac{4m^2-2.8t}{4m^2-t}\left[\frac{1}{1-t/0.71}\right]^2
\label{F1}
\end{equation}
where $m$ is the mass of the proton. When using 
this form factor, 
the pomeron flux 
is referred to as the Donnachie-Landshoff (DL) flux\footnote{
The factor $K$ in the DL flux is  
$K_{DL}={(3\beta_{\pom qq})^2}/{4\pi^2}$,
where $\beta_{\pom qq}$ is the pomeron-quark coupling.}.
Note that at small-$t$ $F_1^2(t)$ can be approximated with an exponential 
expression whose slope parameter, $b(t)=\frac{d}{dt}\ln F_1^2(t)$, is 
4.6 GeV$^{-2}$ at $t\approx -0.04$ GeV$^2$, consistent with the 
slope obtained from elastic scattering at small $t$.

As we discussed in a previous paper~\cite{CMG}, the 
$\sim s^{\textstyle \epsilon}$
dependence of $\sigma_{T}(s)$ violates the
unitarity based Froissart bound, which  states that the
total cross section cannot rise faster than  $\sim \ln^2 s$.
Unitarity is also violated by the $s$-dependence of the ratio
$\sigma_{el}/\sigma_T\sim s^{\textstyle \epsilon}$, 
which eventually exceeds the
black disc bound of one half ($\sigma_{el}\leq \frac{1}{2}\sigma_T$), 
as well as by the $s$-dependence of the integrated diffractive cross section,
which increases with $s$ as  $\sim s^{2{\textstyle \epsilon}}$ 
and therefore grows faster 
than the total cross section. 

For both the elastic and total cross sections,
unitarization can be achieved by eikonalizing
the elastic amplitude~\cite{CMG,GLM}, which takes into account
rescattering effects.
Attempts to introduce rescattering in the diffractive amplitude
by including cuts~\cite{K1,K2} or by eikonalization~\cite{GLM} have met
with moderate success.
Through such efforts, however, it has become clear that these
``shadowing effects" or 
``screening corrections" affect mainly the normalization of the
diffractive amplitude, leaving the form of the $M^2$ dependence almost
unchanged. This feature is clearly present in the data,
as demonstrated by the CDF Collaboration~\cite{CDF} in comparing 
their measured diffractive differential
$\bar p p$ cross sections at $\sqrt{s}=$546 and 1800 GeV  with $pp$
cross sections at $\sqrt{s}=20$ GeV.

Motivated by these
theoretical results  and by the trend observed in the data,
a phenomenological approach to unitarization of the diffractive amplitude
was proposed~\cite{R} based on ``renormalizing" the pomeron flux 
by requiring its integral over all $\xi$ and $t$ to saturate at unity.
Such a renormalization, which
corresponds to {\em a maximum} 
of one pomeron per proton, leads to interpreting the
pomeron flux as a probability density simply describing the $\xi$ and $t$
distributions of the exchanged pomeron in a diffractive process
(see details in section 3).

In this paper, we show that the hypothesis of flux renormalization 
provides a good description not only of the s-dependence of the total 
integrated SD cross section, as was already shown in~\cite{R}, 
but also of the differential $M^2$ (or $\xi$) 
and $t$ distributions. Specifically, we show that 
for $M^2>5$ GeV$^2$ (above the resonance 
region) and $\xi<0.1$ (the coherence region~\cite{KG}),
all available data for $p+p(\bar p)\rightarrow X+p(\bar p)$ at small $t$ 
can be described by a
renormalized triple-pomeron exchange amplitude, 
plus a non-diffractive contribution
from a ``reggeized" pion exchange amplitude, 
whose normalization is fixed at 
the value determined from charge exchange experiments, $pp\rightarrow Xn$. 
A good fit to the data is obtained using only {\em one free parameter},
namely the triple-pomeron coupling, $g(0)$.

We also show that the $t=0$ cross section at small $\xi$ 
displays a striking scaling behavior, namely   
$d^2\sigma/dM^2dt|_{t=0}\approx C/(M^{2})^{\alpha_{\pom}(0)}$, where the 
coefficient $C$ is $s$-independent 
over six orders of magnitude.
In contrast, the $\sim s^{2{\textstyle \epsilon}}$ 
dependence expected from the 
standard triple-pomeron amplitude represents an increase of 
a factor of $6.5$ between $\sqrt s=20$ and 1800 GeV. 
This scaling behavior is predicted by the 
renormalized flux 
hypothesis and provides a stringent and successful test of its validity.

In section 2 we present and discuss the data we use 
in this paper; in section 3 we describe our phenomenological approach 
in fitting the data using the pomeron
flux renormalization and pion exchange models;  
in section 4 we present the results of our fits to data;  
in section 5 we present the case for a scaling law in diffraction; 
and in section 6 we make some concluding remarks on factorization and 
scaling in soft diffraction. 

\section{Data}
The data we use are from fixed target $pp$ 
experiments~\cite{COOL,SCHAMB}, 
from ISR $pp$ experiments~\cite{ALBROW,ARMITAGE}, 
from $S\bar ppS$ Collider $\bar pp$ experiments~\cite{UA4}, and
from $\bar pp$ experiments at the Tevatron Collider~\cite{CDF,E710}. 
Below, we discuss some aspects of the Tevatron Collider data reported by 
CDF~\cite{CDF}.

\subsection{The CDF data}
The CDF Collaboration reported~\cite{CDF} differential cross sections 
for $\bar pp\rightarrow \bar pX$ in the region of $\xi<0.15$ and 
$|t|<\sim 0.15$ GeV$^2$ 
at $\sqrt{s}=546$ and 1800 GeV. The experiment was performed by 
measuring the momentum of the recoil antiproton using a roman pot spectrometer.
No tables containing data points are given in the CDF publication. 
The data are presented in two figures (figures 13 and 14 in~\cite{CDF}), 
which are reproduced here as Figs.~1a and 1b, where the 
number of events is shown as a function of $x_F$ ($x_F=1-\xi$) of the 
recoil antiproton, rather than as a funcion of the antiproton momentum
(as was done in~\cite{CDF}). 
The histogram superimposed on the data in each figure 
is the CDF  fit to the data generated by a Monte Carlo (MC) simulation. 
As an input to the simulation, the following formula was used:
\begin{equation}
\frac{d^2\sigma}{d\xi dt}
=\frac{1}{2}\left[\frac{D}{\xi^{1+\epsilon}}
e^{\textstyle (b_0-2\alpha'\ln \xi)t} 
+I\xi^{\textstyle \gamma}e^{\textstyle b't}\right]
\label{CDFfit}
\end{equation}
The first term in this equation is the triple-pomeron term of 
Eq.~\ref{diffractive}. The second term was introduced 
to account for the non-diffractive background. A connection to 
Regge theory may be made by observing that $\gamma=1$ (0) corresponds 
to pion (reggeon) exchange with a Regge trajectory of intercept 
$\alpha(0)=0$ (0.5) 
(see section 3).  
The factor 
of $\frac{1}{2}$ does not appear in reference~\cite{CDF} and is introduced 
here to account for the fact that we refer to the cross section 
for $\bar pp\rightarrow \bar pX$ and do not include that for 
$pp\rightarrow pX$, as was done by CDF.
The CDF MC simulation took into account the detector acceptance and the 
momentum resolution of the spectrometer. The slope of the pomeron trajectory, 
$\alpha'$, was kept fixed at the value $\alpha'=0.25$ GeV$^{-2}$. The 
values of the 
remaining parameters, as determined from the CDF fits to the data, are 
listed in Table~1, where we include the 
values for the momentum resolution, $\sigma_0$, at $\sqrt{s}=546$ and 
1800 GeV. 

\subsubsection{Acceptance corrected $\xi$-distributions}
Using the information provided in the CDF publication, 
we mapped Figs.~1a and 1b into Figs.~2a and 2b, respectively, in which the 
data are corrected for detector acceptance.  The acceptance was 
obtained from Fig.~2 of reference~\cite{CDF}. The results are 
presented as cross sections, rather than as events, versus $x$. 
The normalization was determined by comparing the data points 
with the CDF MC fits. The number of events corresponding to   
each $x$-bin of the MC histograms in Figs.~1a and 1b 
was converted to an absolute cross section by convoluting the analytic
CDF formula for the differential cross section with the $t$-acceptance 
function and with
the Gaussian $\xi$-resolution function using a normalization that 
reproduces the MC histogram.
The curves in the new figures represent 
Eq.~\ref{CDFfit} convoluted with a Gaussian resolution function of $\xi$,
whose width was determined from the momentum resolution of the 
spectrometer at each energy. 
Specifically, these curves are calculated using the expression~\cite{xi_min}
\begin{equation}
\frac{d\sigma}{d\xi}=\int_{t=0}^{-\infty}\frac{d^2\sigma}{d\xi dt}dt\,;\;\;\;
\frac{d^2\sigma}{d\xi dt}
=\int_{\xi'=1.4/s}^{1}\,\frac{d^2\sigma}{d\xi' dt}\,g(\xi ',\xi)\,d\xi'
\label{FIT-GAUSS}
\end{equation}
where $\frac{d^2\sigma}{d\xi' dt}$ is given by Eq.~\ref{CDFfit} (with 
$\xi\rightarrow \xi'$)
and $g(\xi ',\xi)$ is the Gaussian resolution function given by
\begin{equation}
g(\xi ',\xi)={1 \over \sqrt{2\,\pi}\,\sigma_{0}}\,e^{-(\xi '- \xi)^{2}/2\,
\sigma_{0}^{2}}.
\label{CDF-RESOL}
\end{equation}

As seen in Figs.~2a and 2b, 
expression (\ref{FIT-GAUSS}) provides an excellent fit to the 
acceptance-corrected differential
cross sections, including the unphysical region of negative
$\xi$ values. Thus, 
once the detector experimental resolution is accounted for, the low $\xi$
(or equivalently, the low $M^{2}$) cross section is {\em completely compatible} 
with that expected from extrapolating the cross section from the region 
of $0.95<x_F<0.99$ $(0.05>\xi>0.01)$ 
into the resolution dominated very low-$\xi$ region  
using the triple-pomeron differential cross section {\em shape}.
This behavior rules out the hypothesis of low $\xi$ (low $M^{2}$) 
suppression suggested by
some authors~\cite{PSSX,PSDIS,TAN,ES98}. 

\subsubsection{Cross sections at $t=-0.05$ GeV$^2$}
As can be seen in Fig.~2 of reference~\cite{CDF}, the CDF data 
in the triple-pomeron dominated region of $\xi<0.05$ are concentrated
at low $t$-values, namely $|t|<\sim 0.1\;(0.2)$  GeV$^2$ 
for $\sqrt{s}=546$ (1800) GeV. 
Therefore, direct comparison of the CDF data with other 
experiments should be made for $t$-values 
within these  regions of $t$. Since the CDF paper does not report 
$\xi$-distributions at a fixed value of $t$ in the form of a table,
we extracted such a table for $t=-0.05$ GeV$^2$ from the information given in 
the CDF paper.  The value of $t=-0.05$ GeV$^2$ was chosen 
in order to allow direct comparison of  
the CDF data with the data of reference~\cite{COOL}, for which 
$\xi$-distributions have been published in table form for $t=-0.05$ GeV$^2$
and $\sqrt{s}=14$ and 20 GeV (see Tables~2 and~3). 
The $t=-0.05$ GeV$^2$ CDF points were evaluated
from the data in Figs.~2a and 2b, which represent cross sections 
integrated over $t$, by scaling the cross section at each point in $\xi$
by the ratio 
\begin{equation}
R(\xi)=\frac{d^2\sigma/d\xi dt|_{t=-0.05}}{d\sigma/d\xi},
\label{ratio}
\end{equation}
which was calculated using Eq.~\ref{FIT-GAUSS}.
Figures~3a and 3b display the $t=-0.05$ GeV$^2$ data points grouped into
$\xi$-bins of approximately equal width in a logarithmic scale. 
Figures~3c and 3d display in a linear $\xi$-scale the data for $\xi<0.01$, 
including the 
unphysical region of negative $\xi$-values.   
The horizontal ``error bars" 
represent bin widths. The values of the points plotted in Figs.~3a-d 
are listed in 
Tables~4 and~5. The solid (dashed) curves in the figures represent the CDF fits 
without (with) the convoluted $\xi$-resolution function, 
calculated using Eq.~\ref{CDFfit} (Eq.~\ref{FIT-GAUSS}).
For $\sqrt{s}=546$ (1800) GeV, the effect of the detector resolution 
becomes important for $\xi<0.005$ (0.003).  Immediately 
below these values, the data lie 
higher than the extrapolation of the solid-line fits from the 
larger $\xi$-values (see Figs.~3a,b).
This effect is completely accounted for by the smearing effect of the 
$\xi$-resolution, which also 
accounts for the values of the cross sections in the unphysical negative
$\xi$-regions, as seen in Figs.~3c and 3d 
(for exact numerical comparisons see Tables~4,~5). The effect 
of the resolution on the measured cross sections is quite substantial  
at low $\xi$ and therefore must be taken into consideration  when 
comparing the low-$\xi$ CDF data 
with predictions of unitarization models based on 
low-$\xi$ suppression of the diffractive cross
section~\cite{PSSX,PSDIS,TAN,ES98}.

\subsubsection{$t$-dependence}
We now return to the question of the slopes $b_0$ of the $t$-distributions
(see Table~1).
Theoretically,  the value of $b_0$ for $pp\rightarrow pX$ should be 
the same at all energies and equal to one half of the 
corresponding value  
for $pp\rightarrow pp$ (see Eqs.~\ref{elastic} and~\ref{diffractive}). 
Experimentally, $\frac{1}{2}b_{0,\,el}=4.6$ GeV$^{-2}$~\cite{R}.
The best-fit CDF slope values are $b_0=7.7\pm 0.6$ ($4.2\pm 0.5$) GeV$^{-2}$ 
for $\sqrt{s}=546$ (1800) GeV.  The 1800 GeV value is close to 4.6,
within error, but the 546 GeV slope is significantly larger than 4.6 
GeV$^{-2}$. 
The discrepancy between the slope value measured by CDF at $\sqrt s=546$ GeV 
and the expected 
value of $b_0=4.6$ GeV$^{-2}$ may be explained by the 
very short $t$-range of the experimental measurement.
In the region of low $\xi$, 
where pomeron exchange is dominant, the detector had reasonable 
acceptance only within the region $0.03<|t|<0.1$ GeV$^2$.
Thus, the slope could not be measured accurately.
The quoted error in the measured slope is  the standard deviation calculated 
keeping all other parameters fixed at their best-fit values. The 
large correlation coefficients~\cite{CDF}
between the error of the CDF best-fit parameter $b_0$ and other 
fit parameters 
indicate that a good fit to the data within the $t$-region of the measurement 
could have been obtained with a different 
value of $b_0$, and correspondingly different value of the other 
parameters, 
subject to the constraint that the integrated cross section over the 
$t$-range of the measurement remain the same. Since $t=-0.05$ GeV$^2$ 
corresponds approximately to the cross-section-weighted mean value of $t$ 
in the region $0.03<{t}<0.1$, the value of the differential 
cross sections at $t=-0.05$ GeV$^2$ is insensitive to a change in $b_0$. 

\subsubsection{Total diffractive cross sections}
At $\sqrt{s}=$546 (1800) GeV, the total integrated cross section  
within the region $0>t>-\infty$ and $(1.5\;\mbox{GeV}^2)/s<\xi<0.05$ 
calculated using 
Eq.~\ref{CDFfit} (multiplied by a factor of 2 to include the cross 
section for $\bar pp\rightarrow Xp$) is 7.28 (8.73) mb. 

\section{Phenomenological approach}
In the framework of Regge theory~\cite{COLLINS},
the cross section for $pp\rightarrow pX$ in the region of large $s/M_X^{2}$
can be expressed as a sum of contributions from exchanges of 
reggeons $i$, $j$ and $k$ (see Fig.~4),
\begin{equation}
\frac{d^2\sigma_{sd}}{dM_X^2\,dt}\,=\,\frac{s_0}{s^{2}}
\sum_{i,j,k} G_{ijk}(t)
\left(\frac{s}{M_X^2}\right)^{\alpha_{i}(t)+\alpha_{j}(t)}
\left(\frac{M_X^2}{s_{0}}\right)^{\alpha_{k}(0)}cos[\phi_i(t)-\phi_j(t)]
\label{triple-regge}
\end{equation}
with
\begin{equation}
G_{ijk}(t)=\frac{1}{16\pi}\,\beta_{i pp}(t)\,\beta_{j pp}(t)
\,\beta_{k pp}(0)\,g_{ijk}(t),
\label{Gijk}\end{equation}
where $\alpha_{i}(t)=\alpha_{i}(0)+\alpha_{i}'t$ is a reggeon trajectory,
$\beta_{ipp}$ is the reggeon coupling to the proton, $g_{ijk}$ is the
``triple-reggeon" coupling and $\phi_i(t)$ is a phase factor 
determined by the signature factor, $\eta_i(t)=\zeta+e^{-i\pi \alpha_i(t)}$, 
where $\zeta=\pm 1$ is the signature of the exchange. 
The signature factors have been expressed as 
$\eta_i(t)=\eta_i^0(t)\,e^{i\phi_i(t)}$ with the moduli $\eta_i^0(t)$
absorbed into the $\beta(t)$ parameters in~(\ref{Gijk}).
For $pp\rightarrow pX$ reggeons $i$ and $j$ must have the same signature, 
so that $\phi_i(t)-\phi_j(t)=\frac{\pi}{2}[\alpha_i(t)-\alpha_j(t)]$.
As mentioned in section 1, 
the energy scale $s_0$ is not determined by the theory and is usually set 
to 1 GeV$^2$. The lack of theoretical 
input about the value of $s_0$ introduces an uncertainty in the pomeron flux 
normalization, which is resolved in the renormalized pomeron flux model 
(see discussion below).

Table~6 displays the $s$ and $\xi$, or $M_X^2$, dependence of the
contributions to the $pp\rightarrow pX$ cross section at $t=0$ 
coming from various combinations of exchanged reggeons.
Three Regge trajectories  are considered:
the pomeron, $\pom$, with $\alpha_{\pom}(0)=1+\epsilon$, the reggeon, $R$, 
with $\alpha_R=0.5$, and the pion, $\pi$, with $\alpha_{\pi}=0$.
In fitting elastic and total cross sections, Covolan, Montanha and 
Goulianos~\cite{CMG} use two reggeon trajectories, one for the f/a 
family with $\alpha_{f/a}(0)=0.68$ and the other for the $\rho/\omega$ family 
with $\alpha_{\rho/\omega}(0)=0.46$; Donnachie and Landshoff~\cite{DLT}
use one ``effective" trajectory with $\alpha^{eff}_{R}(0)=0.55$. 
For simplicity and clarity in presentation, we consider in Table~6 one 
reggeon trajectory with $\alpha_R(0)=0.5$.
The terms $\pom \pom \pom$ (triple-pomeron) and $\pom\pom R$ 
correspond to the picture~\cite{IS} 
in which pomerons emitted by one proton interact with the other 
proton to produce the diffractive event. 
The last row in Table~6 shows the predictions of the renormalized 
pomeron flux model, which is discussed below.

\subsection{Standard approach}
The standard approach to diffraction is to 
perform a simultaneous fit to the $pp\rightarrow pX$ differential 
cross sections of all available data at all energies 
using Eq.~\ref{triple-regge}, which is based on factorization. 
In such a fit, the only free parameters 
are the tripple-reggeon couplings, $g_{ijk}(t)$. The reggeon trajectories and 
the couplings $\beta(t)$ are determined from the elastic and 
total $pp$ cross sections~\cite{CMG}, and
the coupling $\beta_{\pi pp}(t)$ is obtained from 
the coupling $\beta_{\pi np}(t)$ measured in 
the charge exchange reaction $pp\rightarrow nX$; 
using isotopic spin symmetry, 
$\beta_{\pi pp}(t)=\frac{1}{2}\beta_{\pi np}(t)$.
\subsubsection{Factorization}
Equation~\ref{triple-regge} is based on factorization. 
A ``global" fit of this form to all available data  
was performed by R. D. Field and G. C. Fox in 1974~\cite{FIELD-AND-FOX}.
However, the data available at that time could not constrain the 
fit well enough to test the triple-reggeon phenomenology, 
let alone determine the triple-reggeon 
couplings. By 1983, with more data available from Fermilab fixed 
target and ISR experiments~\cite{COOL}-\cite{ARMITAGE}, good 
fits to the small-$t$ differential $pp\rightarrow pX$ cross 
sections were obtained using the empirical expression~\cite{KG} 
\begin{equation}
\frac{d^2\sigma}{d\xi\,dt}=\frac{A}{\xi}\cdot e^{\textstyle bt}
+B\cdot \xi\cdot e^{\textstyle b't}
\label{pompi}
\end{equation}
The first term in (\ref{pompi}) 
has the $\xi$-dependence of the $\pom\pom\pom$ 
amplitude with $\alpha_{\pom}(0)=1$ ($\epsilon=0$) 
and the second term has the 
$\xi$-dependence of the $\pi\pi\pom$
amplitude. Note that a 
reggeon-exchange contribution, $RR\pom$, with $\alpha_R(0)=0.5$, 
would have  a flat $\xi$-dependence. 
At the relatively low values of 
$\sqrt{s}$ of the Fermilab fixed target and ISR experiments, the 
$\xi$-range was not large enough for the $b$-slope to be sensitive 
to the variation with $\xi$ expected from Eq.~\ref{triple-regge}, 
namely $b=b_0-2\alpha'\,\ln\xi$, or 
to distinguish between a $1/\xi$ and a $1/\xi^{1+\epsilon}$ 
dependence in the first term of (\ref{pompi}) 
and thereby establish the now well known 
deviation of $\alpha_{\pom}(0)$ from unity. 
Nevertheless, the prominent $\sim 1/\xi$ behavior of the 
cross section at low $\xi$ showed $\pom\pom\pom$ dominance and 
left little room for contributions from other terms, 
as for example from a $\pom\pom R$ term with its sharper 
$\sim 1/\xi^{1+0.5}$
dependence on $\xi$. 
This is illustrated by the fits of Eq.~\ref{pompi} 
to the very precise data for 
$pd\rightarrow Xd$ shown in Fig.~5. The data~\cite{PD1,PD2} are from the 
experiment of the USA-USSR Collaboration at Fermilab using an internal
gas-jet target operated with deuterium. 
The values of the cross sections at $t=-0.035$ GeV$^2$ plotted in Fig.~5  
were obtained either directly 
from the published Tables~\cite{PD2} or by extrapolation from their published 
values at $t=-0.05$ GeV$^2$~\cite{PD1} using the measured slope of the 
$t$-distribution. 
The two sets of data were normalized to the average value of the 
cross section within the $\xi$-region common to both sets of data.
Figures 5a and 5b show fits using a $1/\xi$ and a $1/\xi^{1+\epsilon}$ 
dependence (with $\epsilon=0.104$~\cite{CMG}), respectively. 
Both fits are in good agreement with the data.  

In summary, the agreement of the Fermilab fixed target and ISR 
experimental results with the empirical 
expression~(\ref{pompi}), which is inspired by
the factorization based standard tripple-reggeon phenomenology, shows that:
\begin{itemize}
\item At small $\xi$ the cross section is dominated by the  
$\pom\pom\pom$ amplitude ($\sim 1/\xi$).
\item At larger $\xi$ there is an additional contribution,
which has the form of the $\pi\pi\pom$ amplitude ($\sim \xi$).
\end{itemize}

\subsubsection{Breakdown of factorization}
In 1994, when CDF published the diffractive $pp\rightarrow pX$ cross sections
at $\sqrt{s}=546$ and 1800 GeV~\cite{CDF}, the supercritical 
pomeron trajectory with $\alpha_{\pom}(0)>1$ was already well 
established by fits to total hadronic cross sections~\cite{DLT}. 
Therefore, CDF made fits using Eq.~\ref{CDFfit}, which includes 
two terms: the $\pom\pom\pom$ amplitude (first term)  
and a non-diffractive contribution 
parameterized as $d^2\sigma_{nd}/d\xi dt=I\xi^{\gamma}e^{b't}$. The form 
of the latter was inspired by the empirical 
expression~(\ref{pompi}), and the parameter $\gamma$ was introduced to 
{\em effectively} incorporate possible contributions both from  
$\pi\pi\pom$ ($\gamma=1$) and  $RR\pom$ ($\gamma=0$) amplitudes,
as discussed in section 2.1.

Three important results from the CDF fits to the data should be emphasized:
\begin{itemize}
\item Only the $\pom\pom\pom$ term and a non-diffractive contribution are 
required by the fits.
An upper limit of 15\% was set on 
a possible contribution of a $\pom\pom R$ term to the total 
diffractive cross section at $\sqrt{s}=546$ GeV.
From this result, we derive the following  limit for the ratio, $R$, 
of the coefficients $G_{ijk}(0)$ of the $\pom\pom R$/$\pom\pom\pom$ terms:
\begin{equation}
R\equiv \frac{G_{\pom\pom R}(0)}{G_{\pom\pom\pom}(0)}<0.2
\label{PPP/PPR}
\end{equation}
This limit is 5\% of the value of $R$ used in the fit by 
Erham and Schlein~\cite{PSSX,PSDIS} (see also comments in~\cite{KG-ES}).
\item The parameter $\epsilon$ was 
determined {\em for the first time from the $\xi$-distribution of single 
diffraction dissociation} and was compared to the $\epsilon$ 
obtained from the 
$s$-dependence of the total $\bar pp$ cross section~\cite{CDFT}. 
The CDF results are:
\begin{eqnarray}
\epsilon\;{\rm (from\;total\;cross\;section)}&=0.112\pm 0.013\\
\epsilon\;({\rm from}\;
d\sigma/d\xi; \;\sqrt{s}=546\; {\rm GeV})&=0.121\pm 0.011\\
\epsilon\;({\rm from}\;
d\sigma/d\xi; \;\sqrt{s}=1800\; {\rm GeV})&=0.103\pm 0.017
\end{eqnarray} 
The values obtained from the $d\sigma/d\xi$ distributions are, within the 
quoted uncertainties, consistent with the value determined from the  rise 
of the total 
cross section, as would be expected for pomeron pole dominance.
The weighted average of all three values is~\cite{R} 
\begin{equation}
\epsilon=0.115\pm 0.008.
\label{ECDF}
\end{equation}
\item Using the relation $\xi=M^2/s$, 
the $\pom\pom\pom$ (first) term in Eq.~\ref{CDFfit} can be written 
in terms of $M^2$ as (see also Eq.~\ref{triple-regge}) 
\begin{equation}
\frac{d^2\sigma}{dM^2dt}=\frac{G(0)}{2}\cdot (s/s_0)^{\delta}\cdot
\frac{s_0^{\epsilon}}{(M^2)^{1+\epsilon}}\,e^{(b_0+2\alpha'\ln(s/M^2))t}
\label{CDFfitM}
\end{equation}
where in standard Regge theory $\delta=2\epsilon$.
Treating $\delta$ as a free parameter and performing 
a simultaneous fit to the diffractive cross sections, $\sigma_{sd}$, 
at $\sqrt{s}=20$~\cite{COOL}, 546 and 1800 GeV, CDF obtained 
$\delta=0.030\pm 0.016$. 
\end{itemize}
The last result indicates a breakdown of factorization. 
The observed slower than $\sim~(s/s_0)^{2\epsilon}$ increase of 
the diffractive cross section with energy is necessary to preserve unitarity 
and was predicted in 1986~\cite{K1} by calculations including 
shadowing effects from multiple pomeron exchanges. 
More recent work  based on eikonalization of the 
diffractive amplitude~\cite{GLM} or on the inclusion of cuts~\cite{K2}
shows that shadowing can produce substantial damping of 
the $s$-dependence of the cross section 
but has no appreciable effect on the $M^2$-dependence. 
These predictions are 
in general agreement with 
the conclusions reached by the CDF fits to data. However, 
the damping predicted by the eikonalization model is not sufficient 
to account for the observed $s$-dependence of the total single 
diffraction cross section (see Fig.~6);
the predictions of the 
model based on cuts are in better agreement with the data~\cite{K2}.
 
\subsection{Renormalized pomeron flux approach}
\subsubsection{Triple-pomeron renormalization}
The CDF measurements showed that, just like at Fermilab fixed target and 
ISR energies, the shape of the small $M^2$ (small $\xi$) behavior 
of the diffractive cross section at the Tevatron Collider 
is described well by the $\pom\pom\pom$ amplitude displayed in 
Eq.~\ref{CDFfitM}.
The total diffractive cross section, 
obtained by integrating Eq.~\ref{CDFfitM} over all $t$ 
and over $M^2$ from $M^2_{min}=1.5$ GeV$^2$ to $M^2_{max}=0.1s$,
increases with $s$ as $\sim (s/s_0)^{\delta}$.
For $\delta=2\epsilon$, which is the value for simple pole exchange,
$\sigma_{sd}$ would increase faster than the total $\bar pp$ cross section,
which varies as  
$\sim (s/s_0)^{\epsilon}$, leading to violation of unitarity.
With the experimentally determined  value of $\delta\approx 0$, 
the diffractive cross section remains safely below the total cross section 
as $s$ increases, preserving unitarity.

As we have seen in the previous section, 
introducing shadowing corrections can dampen the increase 
of the diffractive cross section with $s$ and thereby achieve 
the desired unitarization while preserving the $M^2$-dependence 
of the $\pom\pom\pom$ amplitude, 
as required by the data. 
However, the shadowing models do not 
account completely for the $s$-dependence of the data, and 
the two models mentioned above do not predict the same amount of $s$-damping 
of the cross section. In addition, these models are very cumbersome to use in  
calculations of single diffraction, double diffraction and 
double-pomeron exchange processes.

The calculational difficulties 
of unitarity corrections in the standard approach
are overcome  in the ``pomeron flux renormalization" approach  
proposed by Goulianos~\cite{R}.
The renormalized flux 
approach is based on a {\em hypothesis}, rather than on an
actual calculation of unitarity corrections, and therefore can be 
stated as an axiom: 
\vglue 2ex
\fbox{
{\bf The pomeron flux integrated over all phase space saturates at unity.}
}
\vglue 2ex
The standard pomeron flux is displayed in Eq.~\ref{flux}.
Using $F^2(t)=e^{b_0t}$, the integral of the standard flux,
\begin{equation}
N(s)=\displaystyle{\int_{\xi_{min}}^{\xi_{max}}}
\displaystyle{\int_{-\infty}^{0}}f_{\pom/p}(\xi,t)d\xi dt, 
\label{fluxI}
\end{equation}
is given by  
\begin{equation}
N(s)=
\displaystyle{K\,{e^{-r} \over 2\alpha'}\,[\,E_{i}\,(r - 2\epsilon
\ln{\xi_{min}}})-E_{i}\,(r - 2\epsilon \ln{\xi_{max}})],
\label{fluxII}
\end{equation}
where $Ei(x)$ is the exponential integral function\footnote{
$E_i(x)=\gamma+
\ln x+{\displaystyle \sum_{n=1}^{\infty}}\frac{x^n}{n\;n!}$, where
$\gamma=$0.57721$\ldots$ (Euler's constant).},
$r\equiv b_{0} \epsilon/\alpha'$, $\xi_{min}=M_0^2/s=1.5/s$ is the 
effective diffractive threshold, and $\xi_{max}=0.1$~\cite{R}.

The renormalized pomeron flux, $f_N(\xi,t)$, can now be expressed 
in terms of the standard flux, $f_{\pom/p}(\xi,t)$, as follows\footnote{
For a detailed discussion of the role of the scale parameter $s_0$ 
in determining the value of $s$ for which 
$N(s)=1$ see ~\cite{R}.}:
\begin{equation}
f_N(\xi,t)=\left\{ \begin{array}{ll}
f_{\pom/p}(\xi,t)&\mbox{if }N(s)<1\\
N^{-1}(s)\cdot f_{\pom/p}(\xi,t)&\mbox{if }N(s)>1
\end{array}
\right.
\label{fluxN}
\end{equation}

The renormalized $\pom\pom\pom$ contribution to the differential 
cross section is given by
\begin{equation}
\frac{d^2\sigma_{sd}}{d\xi dt}=
{K\over N(s)}\;
{{e^{-2\alpha ' t\,\ln{\xi}}\,F^2(t)}\over \xi^{1+2\epsilon}}\cdot
\sigma_0^{\pom p}\;(s\xi)^{\epsilon} 
\label{DIF-RENOR}
\end{equation}
or, in terms of $M^2$, by
\begin{equation}
\frac{d^2\sigma_{sd}}{dM^2 dt}=
{{K\,s^{2\epsilon}}\over N(s)}\;
{{e^{-2\alpha ' t\,\ln{(M^2/s)}}\,F^2(t)}\over (M^2)^{1+2\epsilon}}\cdot
\sigma_0^{\pom p}\;(M^2)^{\epsilon}\,.
\label{DIF-RENOR-M}
\end{equation}

In the energy interval of $\sqrt{s}=20$ to $2000$ GeV, the standard 
flux integral varies as $\sim s^{2\epsilon}$ (see Fig.~7). 
Thus, flux renormalization 
approximately cancels the $s$-dependence in Eq.~\ref{DIF-RENOR-M} 
resulting in a slowly rising total diffractive cross section. 
Asymptotically, as $s\rightarrow \infty$, 
the renormalized total diffractive cross section reaches a constant value:
\begin{equation}
\lim_{s \rightarrow \infty}\sigma_{sd}^{N}(s) \,=\,
\lim_{s \rightarrow \infty}{\sigma_{sd}(s) \over N(s)} \,=\,
2\,\sigma_{0}^{\pom p}\,e^{r/2}.
\end{equation}

The s-dependence of the integral of expression~(\ref{DIF-RENOR}) 
over all $t$ and $\xi<0.05$,
multiplied by a factor of 2 to account for both $\bar pp\rightarrow \bar pX$
and $\bar pp\rightarrow Xp$, is compared with experimental data for 
$\sigma_{sd}(\xi<0.05)$ in Fig.~6 (from~\cite{R}). In view of the 
systematic uncertainties in the normalization of different sets of data,
which are of ${\cal{O}}(10\%)$, the agreement is excellent.

\subsubsection{Pion exchange contribution}
The form of the empirical expression (\ref{pompi}) suggests that at high $\xi$
the dominant non-$\pom\pom\pom$ concontribution to the cross section 
comes from pion exchange. In Regge theory, the pion exchange contribution 
has the form
\begin{equation}
{d^2\sigma \over d\xi dt}=f_{\pi/p}(\xi,t)\cdot \sigma^{\pi p}(s\xi)\,,
\label{pion}
\end{equation}
where $f_{\pi/p}(\xi,t)$ is the pion flux and $\sigma^{\pi p}(s\xi)$ 
the $\pi p$ total cross section.

In the ``reggeized" one-pion-exchange model~\cite{FIELD-AND-FOX}, 
the pion flux is given by
\begin{equation}
f_{\pi/p}(\xi,t)={1 \over 4\pi} \, {g_{\pi pp}^{2} \over 4\pi}
{|t| \over (t - m_{\pi}^{2})^{2}}\,G_1^2(t)\,\xi^{1-2\alpha_{\pi}(t)}
\end{equation}
where $g^2_{\pi pp}/4\pi \approx 14.6$~\cite{FIELD-AND-FOX} 
is the on mass-shell coupling, 
$\alpha_{\pi}(t)=0.9\,t$ is the pion trajectory, and $G_1^2(t)$ is 
a form factor introduced to account for off mass-shell corrections.
For $G_1(t)$ we use the expression~(see~\cite{SCHLEIN} and references therein)
\begin{equation}
G(t)={2.3 - m_{\pi}^{2} \over 2.3 - t}
\label{G}
\end{equation}

Since the exchanged pions are not far off-mass shell, we use the 
on-shell $\pi p$ total
cross section~\cite{CMG},
\begin{equation}
\sigma^{\pi p}(mb)={1 \over 2} (\sigma^{\pi^{+}p} + \sigma^{\pi^{-}p})=
10.83\,(s\,\xi)^{0.104} + 27.13\,(s\,\xi)^{-0.32}
\end{equation} 

\subsubsection{A one parameter fit to diffraction}
Motivated by the success of the empirical expression (\ref{pompi}) in 
describing the Fermilab fixed target and ISR data, and by the 
similarity between this expression and the CDF fits to data
at Tevatron energies, we have performed a simultaneous fit to 
single diffraction differential cross sections at all energies 
using the formula
\begin{equation}
{d^2\sigma \over d\xi dt}=f_N(\xi,t)\cdot \sigma^{\pom p}(s\xi)+
f_{\pi/p}(\xi,t)\cdot \sigma^{\pi p}(s\xi)\,,
\label{one-par}
\end{equation}
in which the first term is 
the renormalized triple-pomeron amplitude, Eq.~\ref{DIF-RENOR}, 
and the second term is the pion exchange contribution, Eq.~\ref{pion}. 
Results from our fit, 
in which only the triple-pomeron coupling, $g_{\pom\pom\pom}$, is treated as 
a free parameter, are presented in the next section. 
\section{Results}
In this section, we present the results of fits performed to 
experimental data using Eq.~\ref{one-par}, which has two 
contributions: a renormalized triple-pomeron amplitude and a 
reggeized pion 
exchange term. 
\subsection{Differential cross sections}
The experimental $\xi$-distributions are usually distorted in the low-$\xi$ 
region by the resolution in the measurement of the momentum of 
the recoil $p(\bar p)$. 
We therefore check first how well Eq.~\ref{one-par} reproduces the shapes of 
the differential cross sections of the $pp$ data of E396~\cite{COOL} 
at $\sqrt s$=14 and 20 GeV and of the $\bar pp$ data 
of CDF~\cite{CDF} at $\sqrt s$=546 and 1800 GeV in the regions of $\xi$
not affected by detector resolution. 
Figure 8 shows the cross sections $d^2\sigma_{sd}/d\xi dt$ at
$t=-0.05$ GeV$^2$ for E396 and CDF (data from Tables 2, 3, 4 and 5). 
The solid lines represent the best fit 
to the data at each energy using Eq.~\ref{one-par} with the normalizations 
of the triple-pomeron and pion exchange contributions treated as 
free parameters. The quality of these fits indicates that no reggeon terms 
other than the triple-pomeron and pion exchange terms are needed 
to describe the shapes of the differential $\xi$-distributions. 

Figures 9-10 show the result of a simultaneous fit (solid lines) 
to the $t=-0.05$ GeV$^2$ data
of E396 and CDF using Eq.~\ref{one-par} with only the triple-pomeron
coupling as a free parameter. The overall normalization of the data 
was allowed to vary within $\pm 10$\% to account for possible 
systematic effects in the experimental measurements. The shift in the 
normalization of the data at each energy that resulted in the best fit 
is given in each plot. In Fig.~10 the individual contributions of the 
triple-pomeron and pion exchange terms are shown by dashed curves. 
The fit had a $\chi^2=1.0$ per degree of freedom.

The parameters used in the fit are $\alpha_{\pom}(t)=1.104+0.25t$
and $\beta_{\pom pp}(0)=6.57$ GeV$^{-1}$ (4.1 mb$^{\frac{1}{2}}$)
for the triple-pomeron term, 
and those given in section 3.2.2 for the pion exchange term. 
The fit yielded a triple-pomeron coupling 
$g_{\pom\pom\pom}=1.0\mbox{ GeV}^{-1}\;(0.62\mbox{ mb}^{\frac{1}{2}})$,
which corresponds to $\sigma_0^{\pom p}=2.6\mbox{ mb}$; using the $F_1(t)$ 
form factor yields 
$g_{\pom\pom\pom}=1.1\mbox{ GeV}^{-1}$ $(0.69\mbox{ mb}^{\frac{1}{2}})$
and $\sigma_0^{\pom p}=2.8\mbox{ mb}$.

Figure~11 shows a fit of Eq.~\ref{one-par} to ISR data~\cite{ALBROW}
of $d\sigma/d\xi$ versus $\xi$ at fixed $t$. In this fit, the experimental 
$\xi$-resolution was taken into account by 
convoluting Eq.~\ref{one-par} with the Gaussian resolution function 
(\ref{CDF-RESOL}) using $\sigma_0=0.003$. The parameters used in 
Eq.~\ref{one-par} were those derived above. 
The overall
normalization of the data has an
experimental systematic uncertainty of 15\%~\cite{ALBROW}.
\subsection{Total diffractive cross sections}
In Fig.~12, we compare experimental results for the total diffractive cross 
section within $0\leq -t\leq -\infty$ and $\xi=M_X^2/s\leq 0.05$
with the cross section calculated from the triple-pomeron term 
of Eq.~\ref{one-par}  (solid line) 
using the triple-pomeron couplig evaluated from 
our fit to the differential cross sections. Within this region of $\xi$, 
the expected contribution of the pion exchange term is less than 2\% 
at any given energy.
The data points are from 
references~\cite{CDF,COOL,SCHAMB,ALBROW,ARMITAGE,UA4,E710}. 

There are two 
points that must be kept in mind in comparing data with theory:
\begin{itemize}
\item {\bf Normalization of data sets}\\
The {\em overall} normalization uncertainty in each 
experiment is of ${\cal{O}}(10\%)$.
\item {\bf Corrections applied to data}\\
Deriving the 
total cross section from experimental data invariably involves 
extrapolations in $t$ and $\xi$ from the regions of the measurement to 
regions where no data exist. In making such extrapolations, certain 
assumptions are made about the shape of the $t$-distribution and/or 
the shape of the $\xi$ distribution. Different experiments make different 
assumptions. For example, with the exception of the ISR 
experiments~\cite{ALBROW,ARMITAGE}, all measurements of the experiments 
listed here are at very low-$t$.  In these experiments, an exponential 
form factor of the form $e^{\textstyle b_0t}$ is assumed 
for extrapolating into the high-$t$ region.
The (higher-$t$) ISR data show a
clear deviation from exponential behavior and support the $F_1^2(t)$
form factor.   
Using $F_1^2(t)$ instead of $e^{\textstyle b_0t}$ results in a 
{\em larger} total 
cross section by $\sim 5-10$\%, depending on the value of $s$ (smaller 
correction at higher $s$). The magnitude of the correction depends 
on the $\xi$-region (and through $\xi$ on $s$), 
since the $t$-distribution depends not only on the 
form factor but also on $\xi$ throught the term 
$e^{(-2\alpha'\ln \xi)t}$.\\
Another source of error comes from the fact that the slope of the 
$t$-distribution is usually not measured accurately in experiments 
sensitive only to low-$t$. The discussion in section 2.1.3 of the CDF 
measurement at $\sqrt s=546$ GeV illustrates this point.
\end{itemize}
Table~7 presents the total diffractive cross sections corrected for the 
effects mentioned above. The ISR~\cite{ALBROW,ARMITAGE} and 
$S\bar ppS$~\cite{UA4} cross sections were left unchanged, since
they were calculated taking into account the high-$t$ behavior 
of the differential cross section. The cross sections of 
Refs.~\cite{SCHAMB,E710} 
were multiplied by the ratio of the cross section calculated from the 
$\pom\pom\pom$ expression using the $F_1(t)$ form factor to that 
calculated using the simple exponential form factor. Finally,
the cross sections of Refs.~\cite{CDF,COOL}, for which the data are 
within a limited $t$-region and have no reliable slope parameters, 
were calculated as follows: in each case, we evaluated the integrated cross 
section within the $t-\xi$ region of the experiment using the 
parameters determined by the experiment, and then recalculated this cross 
section using the formula of Eq.~\ref{one-par}, adjusting the normalization 
parameter $D$
to obtain the same value for the integrated cross section over the same 
$t-\xi$ region; this formula was 
then integrated over the region $0<|t|<\infty$ and $1.5/s<\xi<0.05$. 
The corrections to values derived directly from the 
published data are of ${\cal{O}}(10\%)$. In view of the 
systematic uncertainties in the normalization of the various data sets, 
as evidenced by the discrepancies among data from different experiments in 
overlapping $s$-regions, Fig.~12 shows 
excellent agreement between the experimental cross sections and the 
predictions of the one-parameter 
fit of Eq.~\ref{one-par} (using the $F_1(t)$ form factor and 
$\sigma_0^{\pom p}=2.8$ mb).

\section{A scaling law in diffraction}
The renormalization of the pomeron flux to its integral over 
all available phase space may be viewed as 
a scaling law in diffraction, which serves to unitarize the 
triple-pomeron amplitude at the expense of factorization.

The breakdown of factorization is illustrated in Fig.~13, where 
cross sections are plotted as a function of $\xi$ at fixed $t$ for 
$\sqrt s=$ 14 and 20 GeV ($\sqrt {\bar{s_1}}$=17 GeV) 
and $\sqrt s_2$=1800 GeV. 
As noted by CDF~\cite{CDF}, while the 
shapes of the $d\sigma/d\xi$ distributions as $\xi$ decreases tend to 
the shape expected from triple pomeron dominance at both energies, 
the normalization of the $s_2$ points is approximately a factor 
of $(s_2/\bar{s_1})^{\epsilon}=2.6$ lower than that of the $\bar{s_1}$ 
points, instead of being a factor of $(s_2/\bar{s_1})^{\epsilon}$ higher, as 
one would expect from factorization (see factor $s'=s\xi $ in 
Eq.~\ref{diffractive}).  

This particular way in which factorization 
breaks down implies that the $d^2\sigma/dM_X^2dt|_{t=0}$ distribution is 
approximately independent of $s$, and therefore {\em scales} with $s$, 
in contrast to the $s^{2\epsilon}$ 
behavior expected from factorization. Figure~14 shows the differential 
cross sections as a function of $M_X^2$ at $t=-0.05$ GeV$^2$ 
$\sqrt s$=14, 20, 546 and 1800 GeV within $\xi$ 
regions not including the resonance region of $M_X^2<5$ GeV$^2$ 
($\sqrt s$=14 and 20 GeV) and not
affected by the detector resolution (for $\sqrt s$=546 and 1800 GeV). 
These cross sections 
are also shown in Fig.~15 for regions of $\xi$ low enough not to be 
affected by the non-pomeron contribution. In Fig.~15, the data are compared 
with a straight line fit of the form $d\sigma/dM_X^2\sim 1/M_X^{1+n}$, 
(solid line) and with the predictions based on factorization (dashed lines).
Clearly, factorization breaks down in a way that gives rise to 
a scaling behavior.

The scaling of the $M_X^2$ distribution is a consequence of the 
pomeron flux renormalization hypothesis, as pointed out in 
section 3.2.1.
Figure~7 shows that the renormalization factor based on flux scaling 
has an approximate $s^{2\epsilon}$ dependence, which cancels the 
$s^{2\epsilon}$ dependence in $d\sigma/dM_X^2$ 
expected from factorization. An exact 
comparison between data and theory is made in Fig.~16, where 
data and predictions of Eq.~\ref{one-par} are shown for t=0. The $t=0$ data 
were obtained from the $t=-0.05$ GeV$^2$ data shown in Fig.~14 
by subtracting the pion exchange 
contribution at $t=-0.05$ GeV$^2$ and calculating 
the $t=0$ cross section assuming a $t$-distribution given by 
$F_1^2(t)e^{(-2\alpha'\ln \xi)t}$.  The excellent 
agreement between 
data and theory over six orders of magnitude justifies our viewing 
the pomeron flux renormalization hypothesis as 
{\em a scaling law in diffraction}.

\section{Conclusions}
We have shown that experimental data on diffractive 
differential cross sections
$d^2\sigma/d\xi dt$ for 
$pp\rightarrow Xp$ and $\bar pp\rightarrow Xp$ at energies from $\sqrt s=$14 
to 1800 GeV, as well as total diffractive 
cross sections (integrated over $\xi$ and $t$), 
are described well by a renormalized triple-pomeron 
amplitude and a reggeized pion exchange contribution, whose 
normalization is kept fixed at the value determined from $pp\rightarrow Xn$. 

The renormalization of the triple-pomeron amplitude consists in dividing the 
pomeron flux of the standard Regge-theory amplitude by its integral 
over all available phase space in $\xi$ and $t$. Such a division 
provides an unambiguous normalization of the pomeron flux, since the 
energy scale factor, $s_0$,  which is implicit in the definition of the  
pomeron proton coupling $\beta_{\pom pp}(0)$ that determines the 
normalization for the standard flux, drops out. Thus, the 
renormalized pomeron flux depends {\em only} 
on the value of $\xi_{min}$ and on the pomeron trajectory, which 
is obtained from fits to elastic and total cross sections.  
Therefore the only {\em free} parameter 
in the renormalized triple-pomeron contribution to soft diffraction 
is the triple-pomeron coupling constant, $g_{\pom\pom\pom}$.
From our fit to the data
we obtained the value $g_{\pom\pom\pom}=1.1$ GeV$^{-1}$.

The scaling of the pomeron flux to its integral  
represents a scaling law in diffraction, which unitarizes 
the diffractive amplitude at the expense of factorization. 
A spectacular graphical representation of this scaling is provided 
by the experimental differential $d\sigma/dM_X^2|_{t=0}$ 
distribution as a function 
of $M_X^2$ for energies from $\sqrt s$=14 to 1800 GeV. This distribution 
shows a clear $\sim 1/(M_X^2)^{1+\epsilon}$ behavior, 
which is independent of $s$ 
over six orders of magnitude, in agreement with expectations from 
the flux renormalization hypothesis and contrary to the 
$\sim s^{2\epsilon}$ behavior expected from the standard theory based 
on factorization.

\section{Acknowledgements}
J. Montanha would like to thank the Brazilian Federal Agency CNPq for
providing him with a fellowship and The Rockefeller University 
for its hospitality.

\begin{table}[htp]
\begin{center}
\caption{CDF fit-parameters from reference~[10].}
\vglue 0.25in
\begin{tabular}{|c|c|c|}
\hline
         &                       &                        \\
         & $\sqrt{s}=546\,\,GeV$ & $\sqrt{s}=1800\,\,GeV$ \\
         &                       &                        \\ \hline
         &                       &                        \\
$D$      & $3.53 \pm 0.35$       & $2.54 \pm 0.43$        \\
$b_{0}$  & $7.7 \pm 0.6$          & $4.2 \pm 0.5$          \\
$\alpha'$ & $0.25 \pm 0.02$      & $0.25 \pm 0.02$        \\
$\epsilon$ & $0.121 \pm 0.011$   & $0.103 \pm 0.017$      \\
$I$      & $537^{+498}_{-280}$    & $162^{+160}_{-85}$       \\
$\gamma$ & $0.71 \pm 0.22$       & $0.1 \pm 0.16$         \\
$b'$     & $10.2 \pm 1.5$        & $7.3 \pm 1.0$          \\
$\sigma_{0}$ & $1.4\,\,10^{-3}$  & $8.9\,\,10^{-4}$       \\
         &                       &                        \\ \hline
\end{tabular}
\end{center}
\label{CDF-PARAM}
\end{table}
\vglue 0.5in
\begin{minipage}[t]{3.1in}
\begin{center}
Table 2: Differential cross sections at 14 GeV and
$|t|=0.05$ GeV$^2$ [11].\\
\vglue 0.25in
\begin{tabular}{|c|c|} \hline 
           &                      \\
$\xi$      & $d^{2}\sigma/d\xi dt$\\
&$(mb/GeV^2)$\\
           &                  \\ \hline
           &                     \\
$0.0160$   &  $282.0\pm 11.8$      \\
$0.0267$   &  $145.6\pm 9.3$       \\
$0.0373$   &  $112.0\pm 8.4$       \\
$0.0480$   &  $100.0\pm 7.7$       \\
$0.0586$   &  $85.8\pm 7.3$       \\
$0.0693$   &  $79.7\pm 7.1$       \\
$0.0800$   &  $69.1\pm 7.6$       \\
$0.0906$   &  $65.4\pm 7.4$       \\
$0.1013$   &  $51.0\pm 7.5$       \\
           &                          \\ \hline
\end{tabular}
\end{center}
\end{minipage}
\begin{minipage}[t]{3.1in}
\begin{center}
Table 3: Differential cross sections at 20 GeV and
$|t|=0.05$ GeV$^{2}$~[11].\\
\vglue 0.25in
\begin{tabular}{|c|c|} \hline 
           &                    \\
$\xi$      & $d^{2}\sigma/d\xi dt$ \\
&$(mb/GeV^2)$\\
           &               \\ \hline
           &                       \\
$0.0160$   &  $233.2\pm 10.9$      \\
$0.0267$   &  $146.7\pm 7.9$       \\
$0.0373$   &  $105.9\pm 7.0$       \\
$0.0480$   &  $78.8\pm 6.2$       \\
$0.0586$   &  $80.7\pm 6.5$       \\
$0.0693$   &  $70.2\pm 6.1$       \\
$0.0800$   &  $57.0\pm 5.8$       \\
$0.0906$   &  $62.5\pm 6.6$       \\
$0.1013$   &  $68.6\pm 7.0$       \\
           &              \\ \hline
\end{tabular}
\end{center}
\end{minipage}
\clearpage
\setcounter{table}{3}
\begin{table}
\begin{center}
\label{T546}
\caption{Differential cross sections at 546 GeV and
$|t|=0.05$ GeV$^{2}$
extracted from the CDF measurements~[10] (see text for
details).}
\vglue 0.25in
\begin{tabular}{|c|c|c|c|} \hline
             &             &    &          \\
$\xi$        & $d^{2}\sigma/d\xi dt$& CDF Fit&Fit$\otimes$Gauss   \\
             &$(mb/GeV^{2})$&$(mb/GeV^{2})$&$(mb/GeV^{2})$\\
             &             &     &         \\ \hline
             &             &      &        \\
$-0.0046$    &  $21.18\pm 4.73$&&16.95       \\
$-0.0027$    &  $539.1\pm 40.3$&&569.1       \\
$-0.0009$    &  $3534.4\pm 124.0$&&3591.8      \\
$ 0.0009 $   &  $4561.3\pm 135.1$&2568.1&4561.3      \\
$ 0.0027 $   &  $1682.9\pm 73.8$&772.5&1618.2       \\
$ 0.0046 $   &  $563.0\pm 38.8$&443.4&536.2       \\
$ 0.0064 $   &  $300.9\pm 30.5$&308.7&329.0       \\
$ 0.0082 $   &  $226.5\pm 25.7$&236.3&244.8       \\
$ 0.0100 $   &  $178.2\pm 22.6$&191.5&195.9       \\
$ 0.0119 $   &  $136.5\pm 20.6$&161.3&163.9       \\
$ 0.0146 $   &  $107.3\pm 13.9$&131.0&132.3       \\
$ 0.0192 $   &  $82.4\pm 11.3$&101.0&101.6       \\
$ 0.0256 $   &  $77.9\pm 11.7$&78.5&78.8       \\
$ 0.0348 $   &  $67.0\pm 11.6$&62.5&62.6       \\
$ 0.0458 $   &  $49.6\pm 9.8$&53.3&53.3        \\
$ 0.0577 $   &  $54.4\pm 10.6$&48.6&48.6       \\
$ 0.0714 $   &  $41.2\pm 7.8$&46.4&46.4        \\
$ 0.0870 $   &  $47.7\pm 7.9$&45.9&45.9        \\
$ 0.109 $    &  $44.5\pm 7.1$&47.0&47.0        \\
             &             &   &           \\ \hline
\end{tabular}
\end{center}
\end{table}

\begin{table}
\begin{center}
\label{T1800}
\caption{Differential cross sections at 1800 GeV and
$|t|=0.05$ GeV$^{2}$
extracted from the CDF measurements~[10] (see text for
details).}
\vglue 0.25in
\begin{tabular}{|c|c|c|c|} \hline 
             &             &         &     \\
$\xi$        & $d^{2}\sigma/d\xi dt$& CDF Fit&Fit$\otimes$Gauss   \\
             &$(mb/GeV^{2})$&$(mb/GeV^{2})$&$(mb/GeV^{2})$\\
             &             &          &    \\ \hline
             &             &           &   \\
$-0.0022$    &  $375.4\pm 48.0$    & &307.8       \\
$-0.0011$    &  $3419.4\pm 182.8$   &&3419.4      \\
$ 0.0000$    &  $8368.6\pm 278.9$  &&8368.6      \\
$ 0.0011$    &  $5646.9\pm 210.4$  &1603.4&5019.4      \\
$ 0.0022$    &  $1311.9\pm 88.4$   &776.6&1311.9       \\
$ 0.0033 $   &  $568.7\pm 66.5$    &513.7&573.6      \\
$ 0.0044 $   &  $403.4\pm 57.7$    &386.1&404.5   \\
$ 0.0055 $   &  $319.6\pm 52.9$    &311.3&320.0      \\
$ 0.0072 $   &  $222.7\pm 35.8$    &243.7&247.4    \\
$ 0.0100 $   &  $196.7\pm 33.8$    &182.9&184.2   \\
$ 0.0139 $   &  $153.6\pm 29.3$    &140.1&140.6   \\
$ 0.0189 $   &  $106.7\pm 22.1$    &112.1&112.3   \\
$ 0.0250 $   &  $84.6\pm 18.8$     &93.8&93.9   \\
$ 0.0322 $   &  $90.2\pm 18.7$     &81.6&81.7      \\
$ 0.0422 $   &  $73.9\pm 13.7$     &72.2&72.3     \\
$ 0.0555 $   &  $55.0\pm 9.4$     &65.3&65.3    \\
$ 0.0717 $   &  $69.9\pm 10.0$     &60.8&60.8       \\
$ 0.0918 $   &  $57.6\pm 7.2$     &57.8&57.8    \\
$ 0.116 $    &  $55.4\pm 6.5$     &55.8&55.8     \\
         &        &             &              \\ \hline
\end{tabular}
\end{center}
\end{table}

\newpage
\begin{table}
\begin{center}
\label{T-TR}
\caption{Triple-reggeon amplitudes for $pp\rightarrow pX$.}
\vglue 0.15in
$$\alpha_{\pom}(0)=1+\epsilon\;\;\;\;\;\alpha_R(0)=0.5\;\;\;\;\;
\alpha_{\pi}(0)=0$$
\vglue 0.15in
{\Large
\begin{tabular}{llll}
Amplitude &$d^2\sigma/d\xi dt|_{t=0}$&$d^2\sigma/dM^2 dt|_{t=0}$&
$\sigma_{SD}^{tot}(s)$\\
\hline\hline
 & & & \\
$(\pom\pom)\pom$&$\sim \frac{s^{\epsilon}}{\xi^{1+\epsilon}}$&
$\sim \frac{s^{2\epsilon}}
{(M^2)^{1+\epsilon}}$&$\sim s^{2\epsilon}$\\
&&&\\
$(\pom\pom)R$&$\sim \frac{1/\sqrt{s}}{\xi^{1.5+2\epsilon}}$&$
\sim \frac{s^{2\epsilon}}
{(M^2)^{1.5+2\epsilon}}$&$\sim s^{2\epsilon}$\\
&&&\\
$(RR)\pom$&$\sim {s^{\epsilon}}{\xi^{\epsilon}}$&$\sim \frac{1}{s}
{(M^2)^{\epsilon}}$&$\sim s^{\epsilon}$\\
&&&\\
$(RR)R$&$\sim \frac{1/\sqrt{s}}{\xi^{0.5}}$&$\sim \frac{1/s}
{(M^2)^{0.5}}$&$\sim 1/\sqrt{s}$\\
&&&\\
$(\pi\pi)\pom$&$\sim s^{\epsilon}\xi^{1+\epsilon}$&$\sim
\frac{1}{s^2}
\,(M^2)^{1+\epsilon}$&$\sim s^{\epsilon}$\\
&&&\\
$(\pi\pi)R$&$\sim (1/\sqrt{s})\xi^{0.5}$&$\sim \frac{1}{s^2}
(M^2)^{0.5}$&$\sim {1}/{\sqrt{s}}$\\
&&&\\
$(\pom R)R$&$\sim \frac{1/\sqrt{s}}{\xi^{1+\epsilon}}$
&$\sim \frac{s^{\epsilon}/\sqrt{s}}
{(M^2)^{1+\epsilon}}$&$\sim s^{\epsilon}/\sqrt{s}$\\
&&&\\
$(\pom R)\pom$&$\sim \frac{s^{\epsilon}}{\xi^{0.5}}$&$\sim
\frac{s^{\epsilon}/\sqrt{s}}
{(M^2)^{0.5}}$&$\sim s^{\epsilon}$\\
&&&\\
\hline
&&&\\
Renormalized:&&&\\
${(\pom\pom)\pom}$
&$\sim \frac{1/s^{\epsilon}}{\xi^{1+\epsilon}}$&$\sim \frac{1}
{(M^2)^{1+\epsilon}}$&$\sim$ constant\\
&&&\\
\hline
\hline
\end{tabular}
}
\end{center}
\end{table}
\newpage
\begin{table}[htb]
\begin{center}
\caption{Total single diffraction cross sections for $\xi \leq 0.05$.
The cross sections of the references marked with $\dagger$  
were derived from the experimental data 
using the procedure outlined in the text.}
\vglue 0.25in
\begin{tabular}{|r|c|r|} \hline
                 &                       &           \\
$\sqrt{s}$ [GeV] & $\sigma_{sd}$ [mb]    & Ref. \\
                 &(both sides)&\\
&                       &           \\ \hline
                 &                       &           \\
$14$ & $3.94 \pm 0.20$ 
& $\dagger$~\cite{COOL} \\
$20$ & $4.46 \pm 0.25$ 
& $\dagger$~\cite{COOL} \\
$16.2$ & $4.87 \pm 0.08$ 
& $\dagger$~\cite{SCHAMB}  \\
$17.6$  &$4.96\pm 0.08$ 
& $\dagger$~\cite{SCHAMB}  \\
$19.1$ & $4.94 \pm 0.08$ 
& $\dagger$~\cite{SCHAMB}  \\
$23.8$ & $5.19 \pm 0.08$ 
& $\dagger$~\cite{SCHAMB}  \\
$27.2$ & $5.42 \pm 0.09$ 
& $\dagger$~\cite{SCHAMB}  \\
$23.4$ & $6.07 \pm 0.17$ & \cite{ALBROW}  \\
$26.9$ & $6.05 \pm 0.22$ & \cite{ALBROW}  \\
$30.5$ & $6.37 \pm 0.15$ & \cite{ALBROW}  \\
$32.3$ & $6.32 \pm 0.22$ & \cite{ALBROW}  \\
$35.2$ & $7.01 \pm 0.28$ & \cite{ALBROW}  \\
$38.3$ & $6.08 \pm 0.29$ & \cite{ALBROW}  \\
$23.3$ & $6.5 \pm 0.2$ & \cite{ARMITAGE}  \\
$27.4$ & $6.3 \pm 0.2$ & \cite{ARMITAGE}  \\
$32.4$ & $6.5 \pm 0.2$ & \cite{ARMITAGE}  \\
$35.5$ & $7.5 \pm 0.5$ & \cite{ARMITAGE}  \\
$38.5$ & $7.3 \pm 0.4$ & \cite{ARMITAGE}  \\
$44.7$ & $7.3 \pm 0.3$ & \cite{ARMITAGE}  \\
$53.7$ & $7.0 \pm 0.3$ & \cite{ARMITAGE}  \\
$62.3$ & $7.5 \pm 0.3$ & \cite{ARMITAGE}  \\
$546$  & $9.4 \pm 0.7$ & \cite{UA4}\\
$1800$ & $8.46 \pm 1.77$ & $\dagger$~\cite{E710} \\
$546$ & $8.34 \pm 0.36$ 
& $\dagger$~\cite{CDF} \\
$1800$ & $9.12 \pm 0.46$ 
& $\dagger$~\cite{CDF} \\
         &        &             \\ \hline
\end{tabular}
\end{center}
\label{TOTAL}
\end{table}
\clearpage
\newpage
\normalsize
\begin{center}
{\bf\LARGE CDF Single Diffractive Data}\\
(uncorrected for acceptance)
\end{center}
\vglue 0.25in

\centerline{\psfig{figure=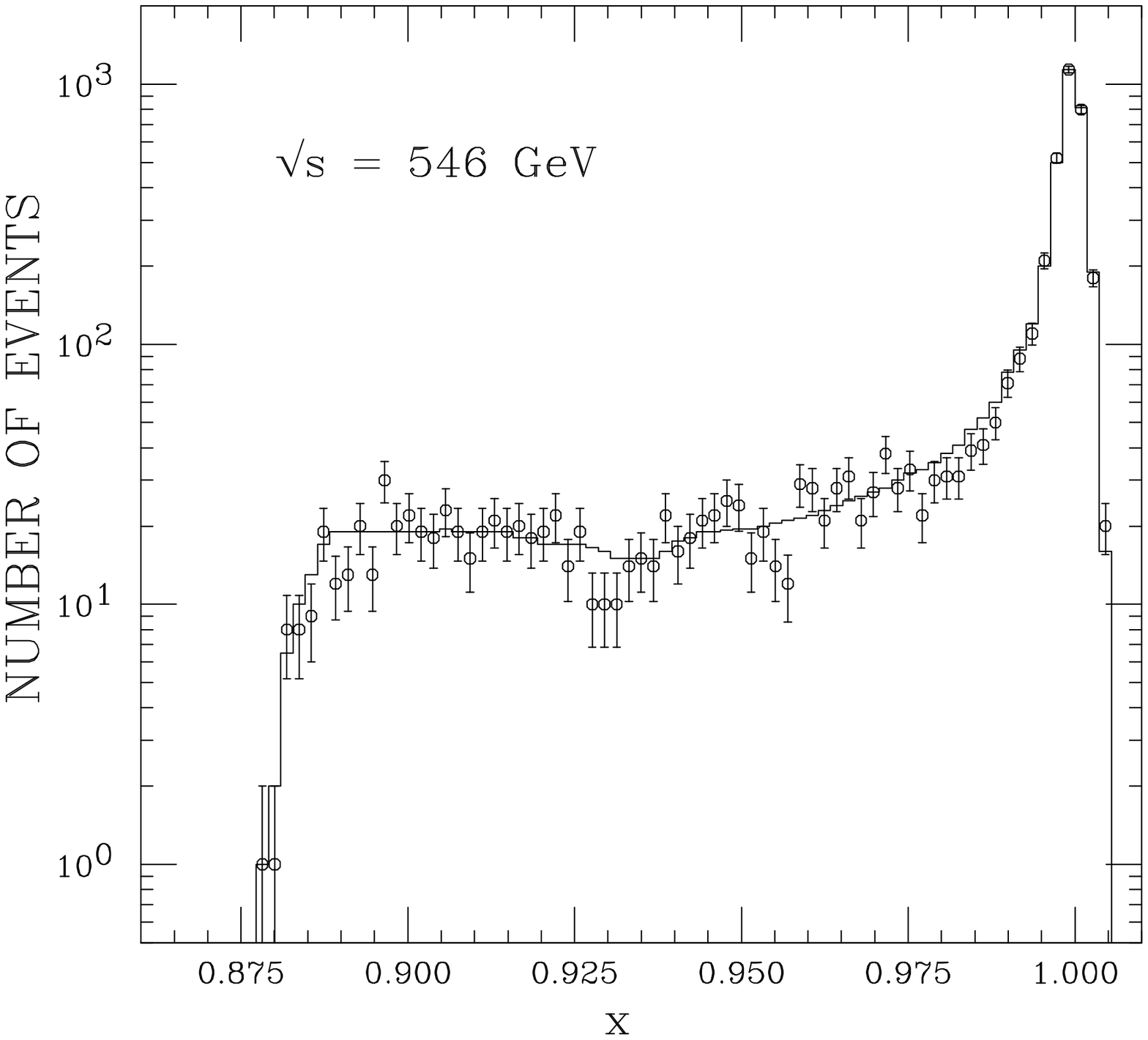,width=3.5in}}
\vglue -0.25in
\centerline{\psfig{figure=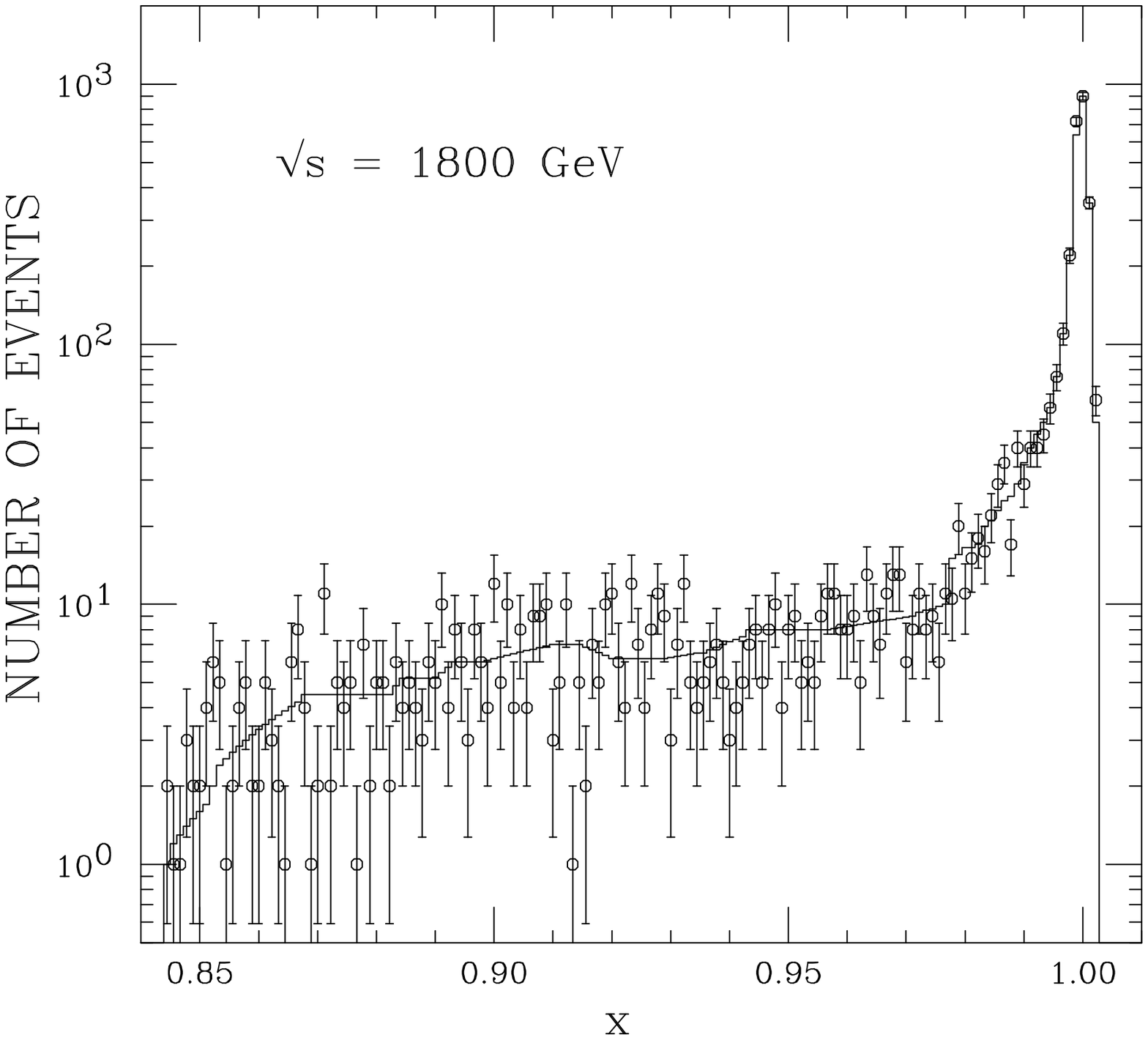,width=3.5in}}
\vglue -0.45in
\centerline{Figures 1a (top) and 1b (bottom)}
\begin{center}
CDF single diffraction data (uncorrected for acceptance): number 
of events versus $x_F$; the solid line histograms are from a Monte Carlo 
simutation using formula (7).
\end{center}
\clearpage
\newpage
\begin{center}
{\bf\LARGE CDF Single Diffractive Cross Sections}\\
(corrected for acceptance)
\end{center}
\vglue 0.25in

\centerline{\psfig{figure=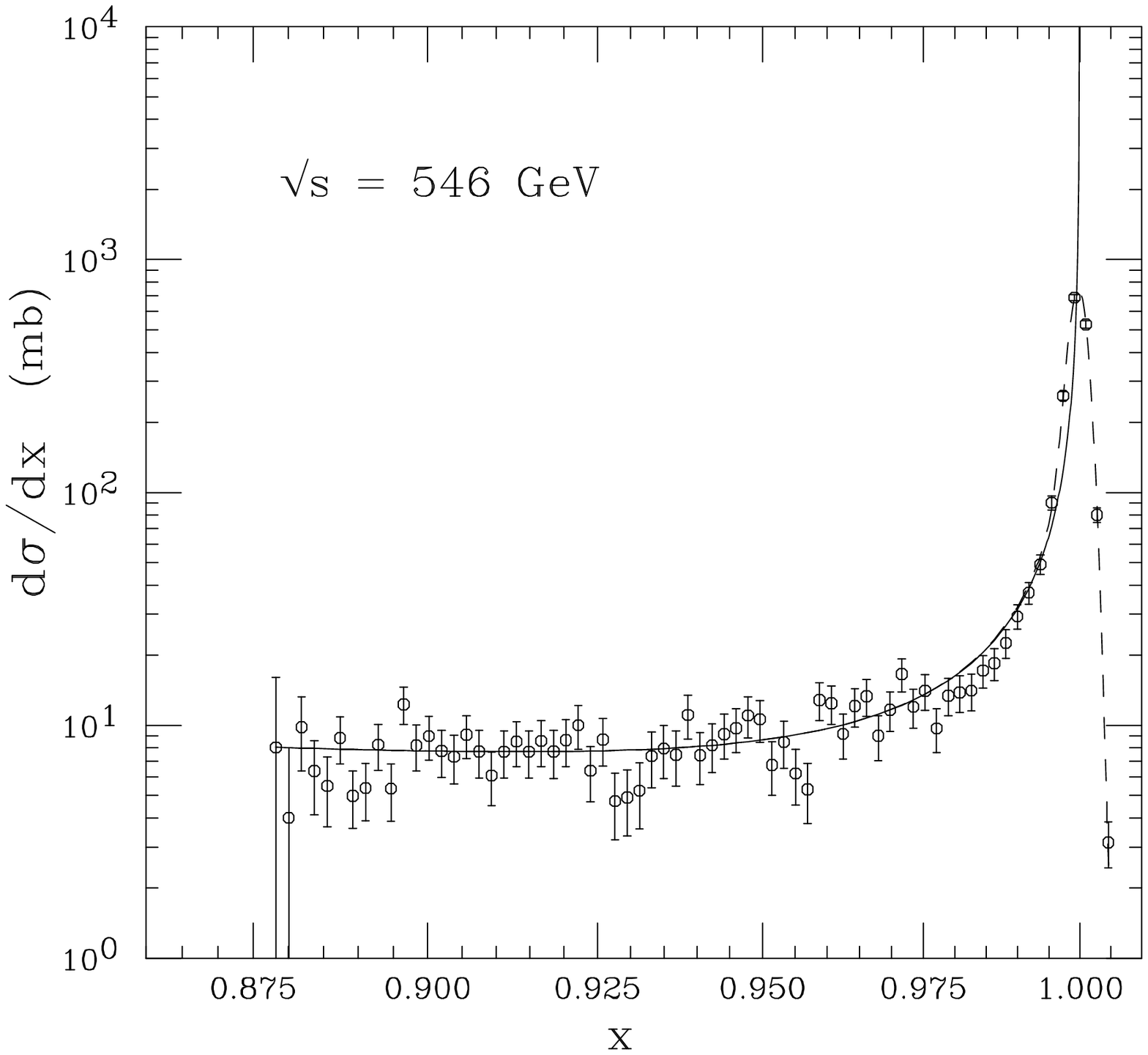,width=3.5in}}
\vglue -0.25in
\centerline{\psfig{figure=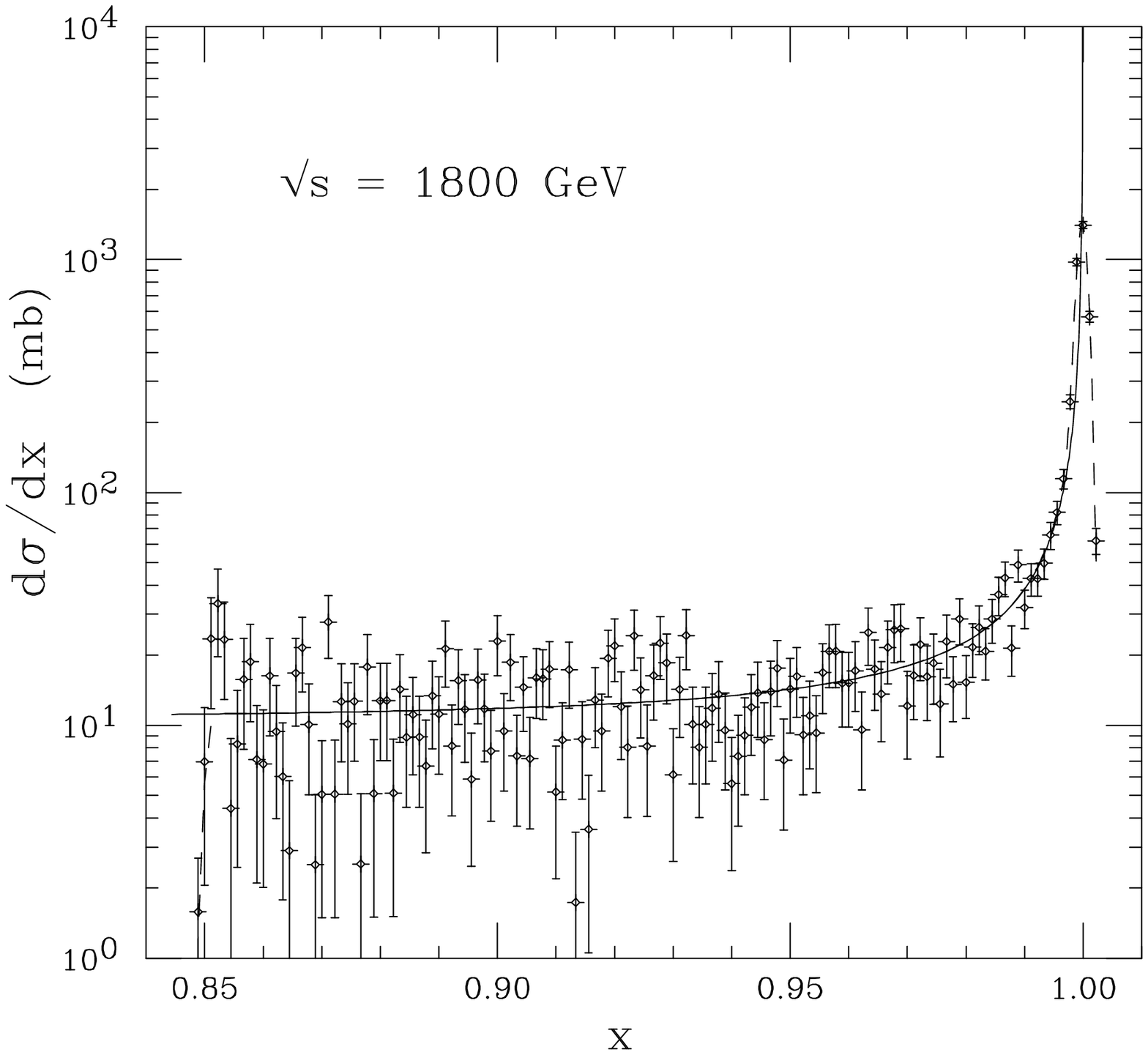,width=3.5in}}
\vglue -0.45in
\centerline{Figures 2a (top) and 2b (bottom)}
\begin{center}
CDF cross sections $d\sigma/dx_F$ (integrated over $t$); 
the solid curves represent 
formula (7) and the dashed curves formula (8).
\end{center}
\clearpage
\newpage
\begin{center}
{\bf\LARGE CDF Single Diffractive Cross Sections}\\
($d\sigma/d\xi$ at $t=-0.05$ GeV$^2$)
\end{center}

\centerline{\psfig{figure=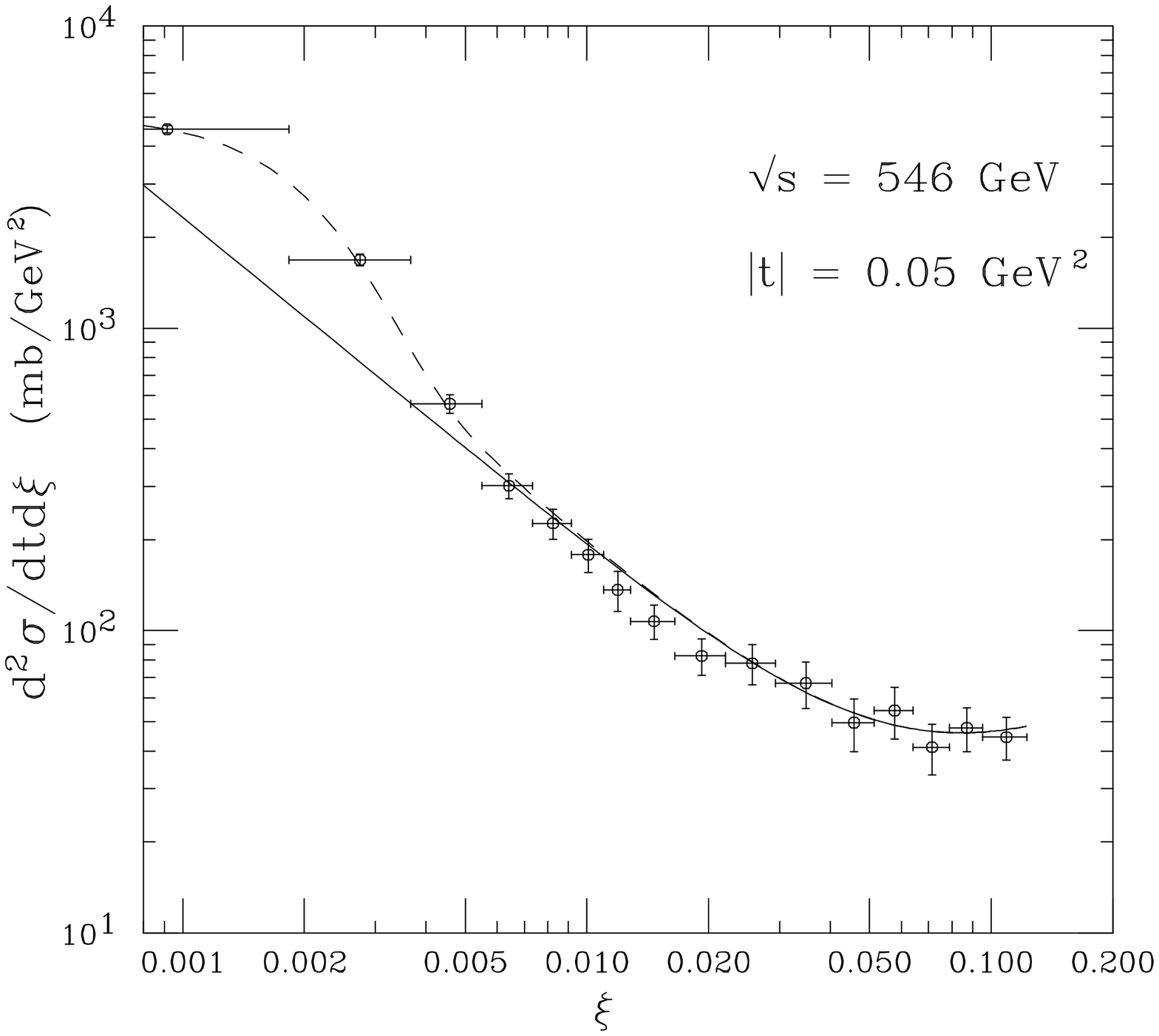,width=3.5in}}
\vglue -0.25in
\centerline{\psfig{figure=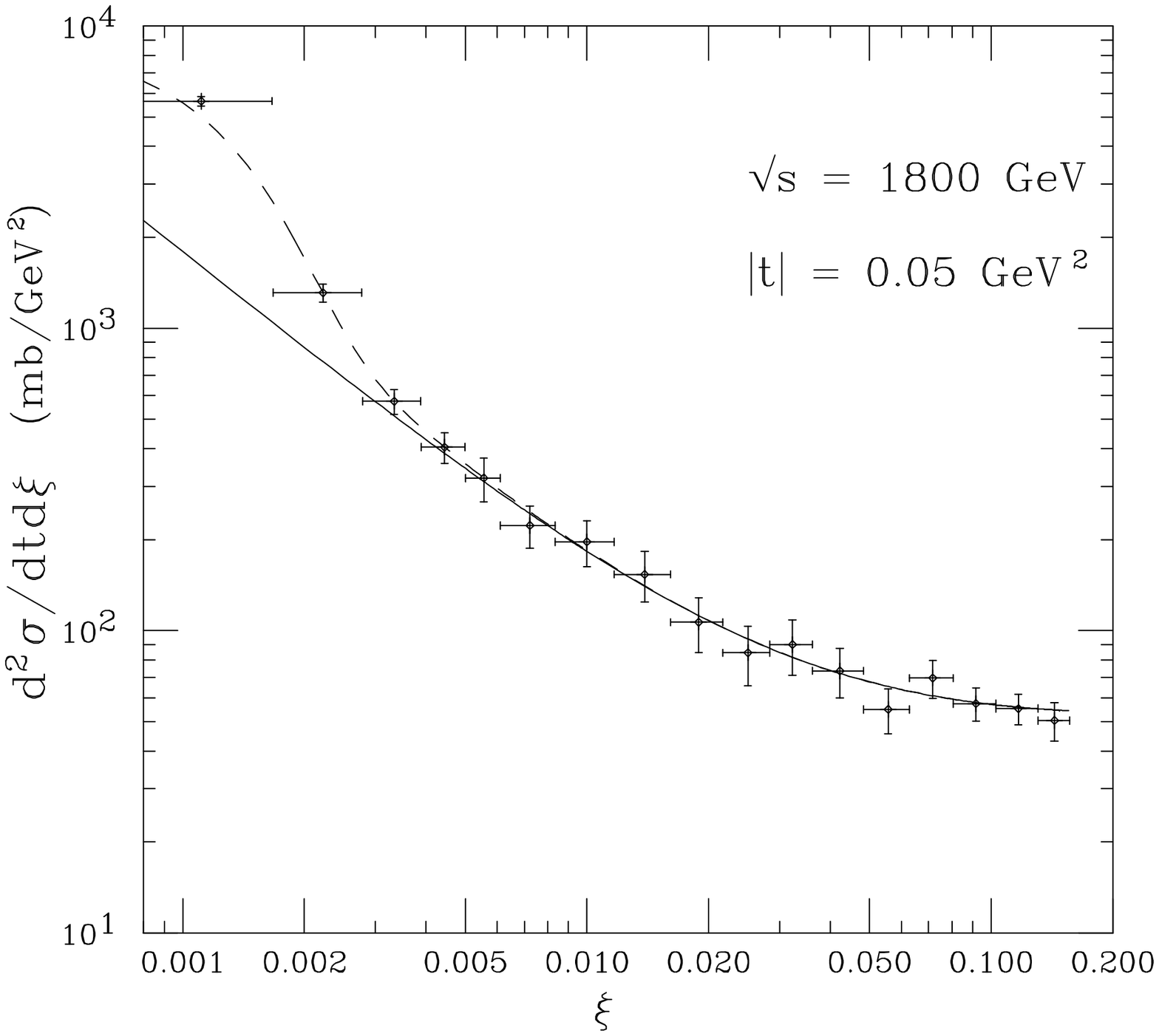,width=3.5in}}
\vglue -0.25in
\centerline{Figures 3a (top) and 3b (bottom)}
\begin{center}
CDF cross sections $d\sigma/d\xi dt$ at $t=-0.05$ GeV $^2$;
the solid curves represent 
formula (7) and the dashed curves formula (8).
\end{center}
\clearpage
\newpage
\begin{center}
{\bf\LARGE CDF Single Diffractive Cross Sections}\\
($d\sigma/d\xi$ at $t=-0.05$ GeV$^2$)
\end{center}

\vglue 0.5cm
\centerline{\psfig{figure=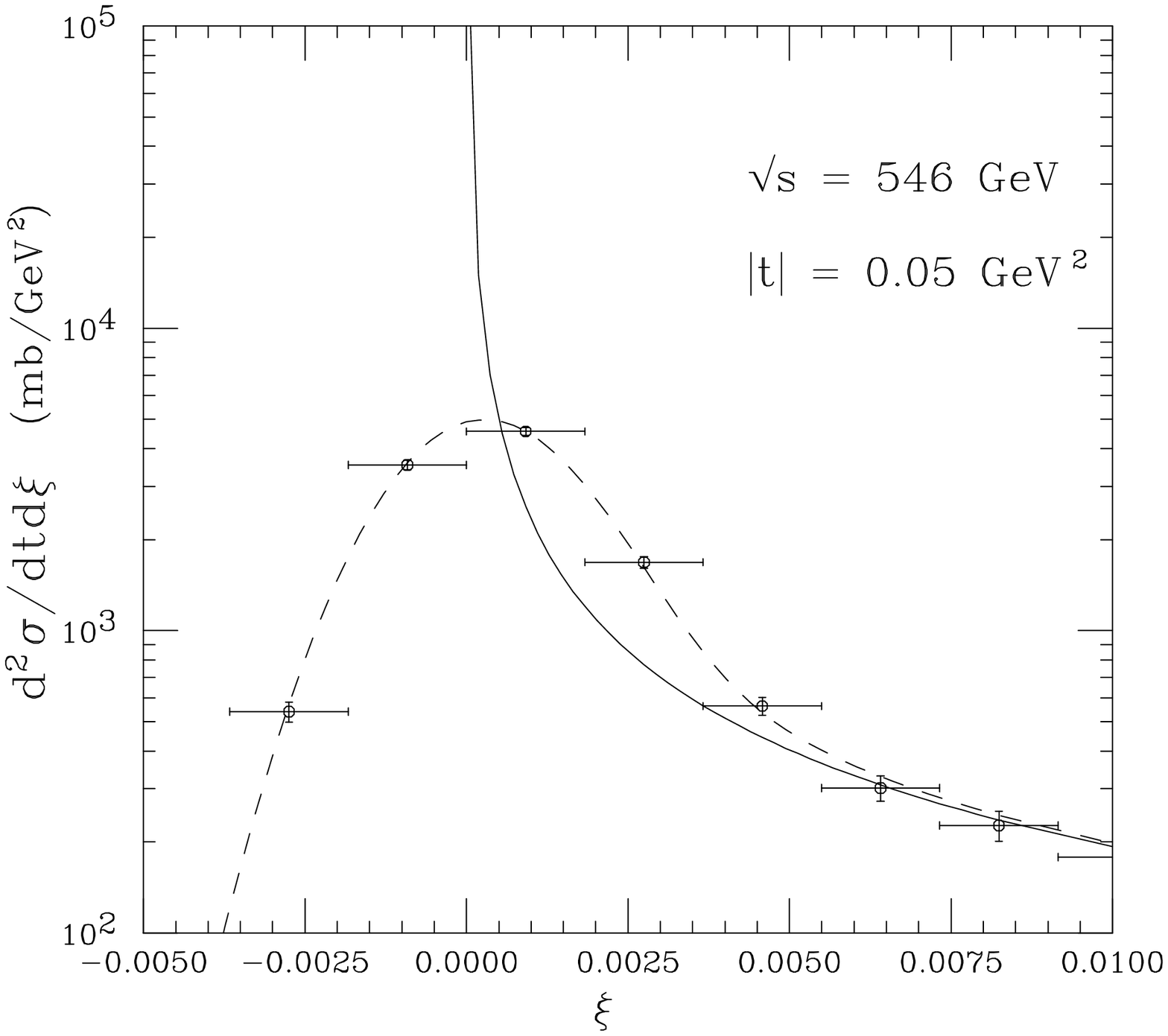,width=3.5in}}
\vglue -0.25in
\centerline{\psfig{figure=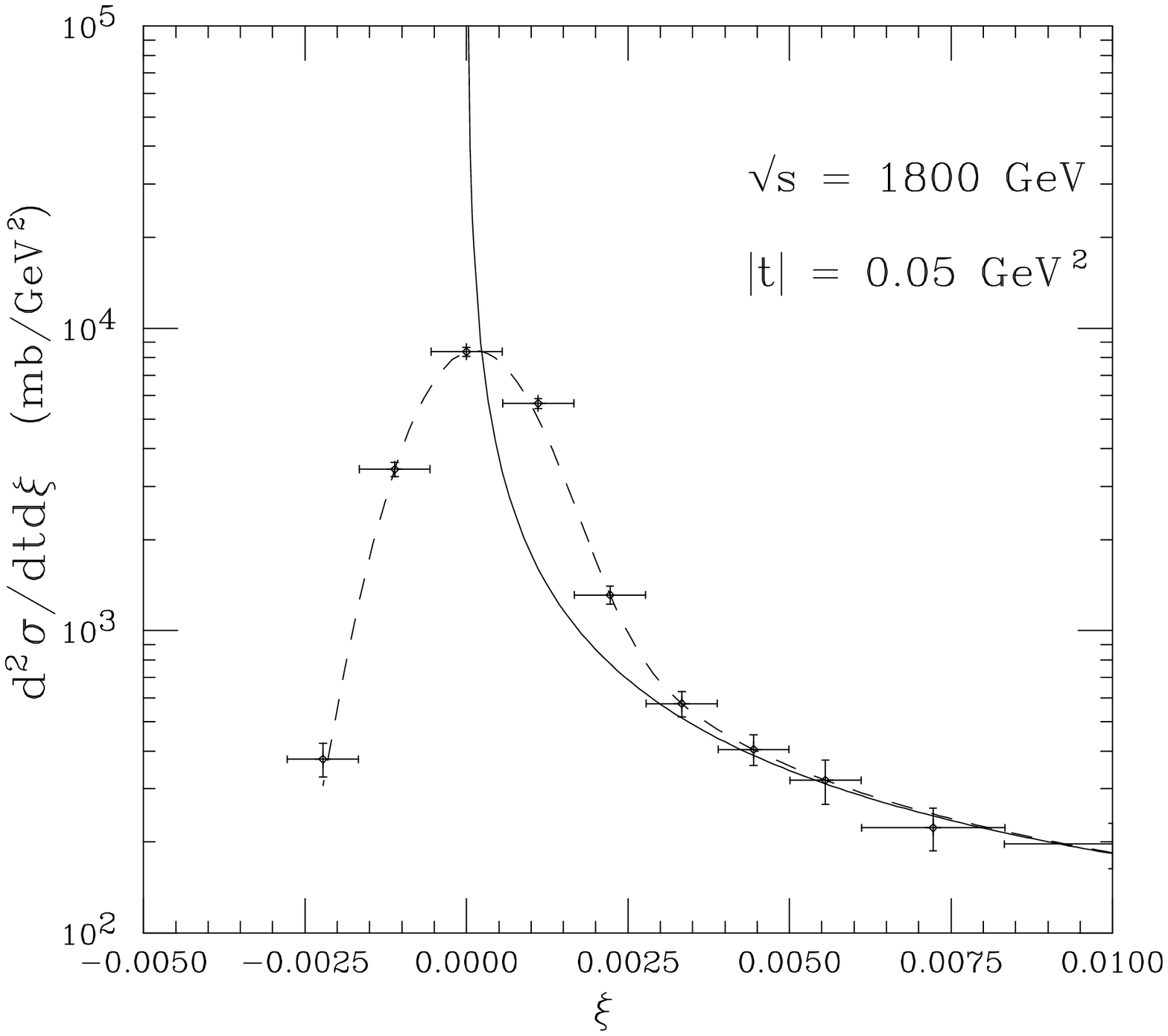,width=3.5in}}
\vglue -0.25in
\centerline{Figures 3c (top) and 3d (bottom)}
\begin{center}
CDF cross sections $d\sigma/d\xi dt$ at $t=-0.05$ GeV $^2$;
the solid curves represent 
formula (7) and the dashed curves formula (8).
\end{center}
\clearpage
\newpage
\vglue 2in
\centerline{\psfig{figure=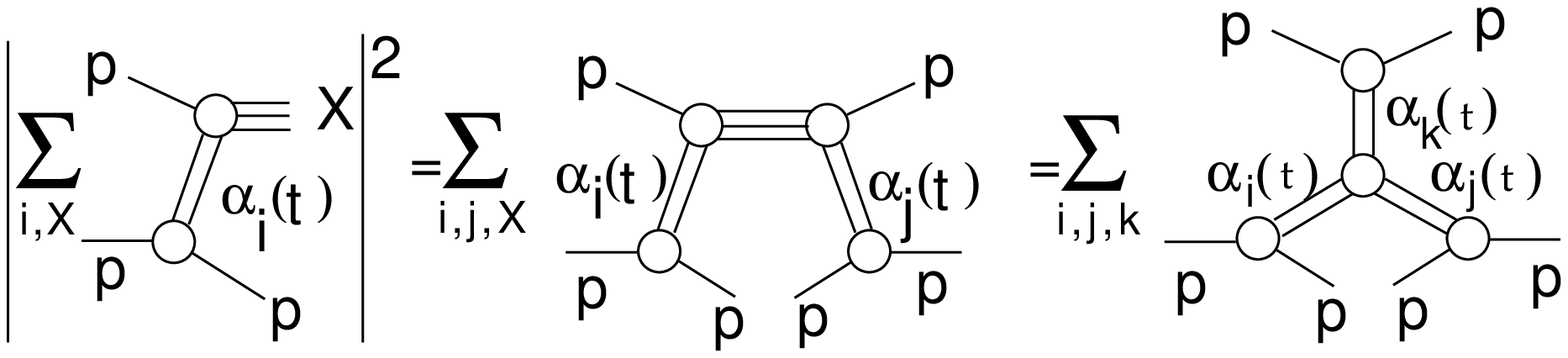,width=6in}}
\vglue -4in
\centerline{Figure 4: Illustration of the triple-reggeon phenomenology.}
\clearpage
\newpage
\begin{center}
{\bf\LARGE Cross Sections for $p+d\rightarrow X+d$\\
($\xi\cdot d\sigma/d\xi$ at $t=-0.035$ GeV$^2$)}
\end{center}
\vglue 1cm
\centerline{\psfig{figure=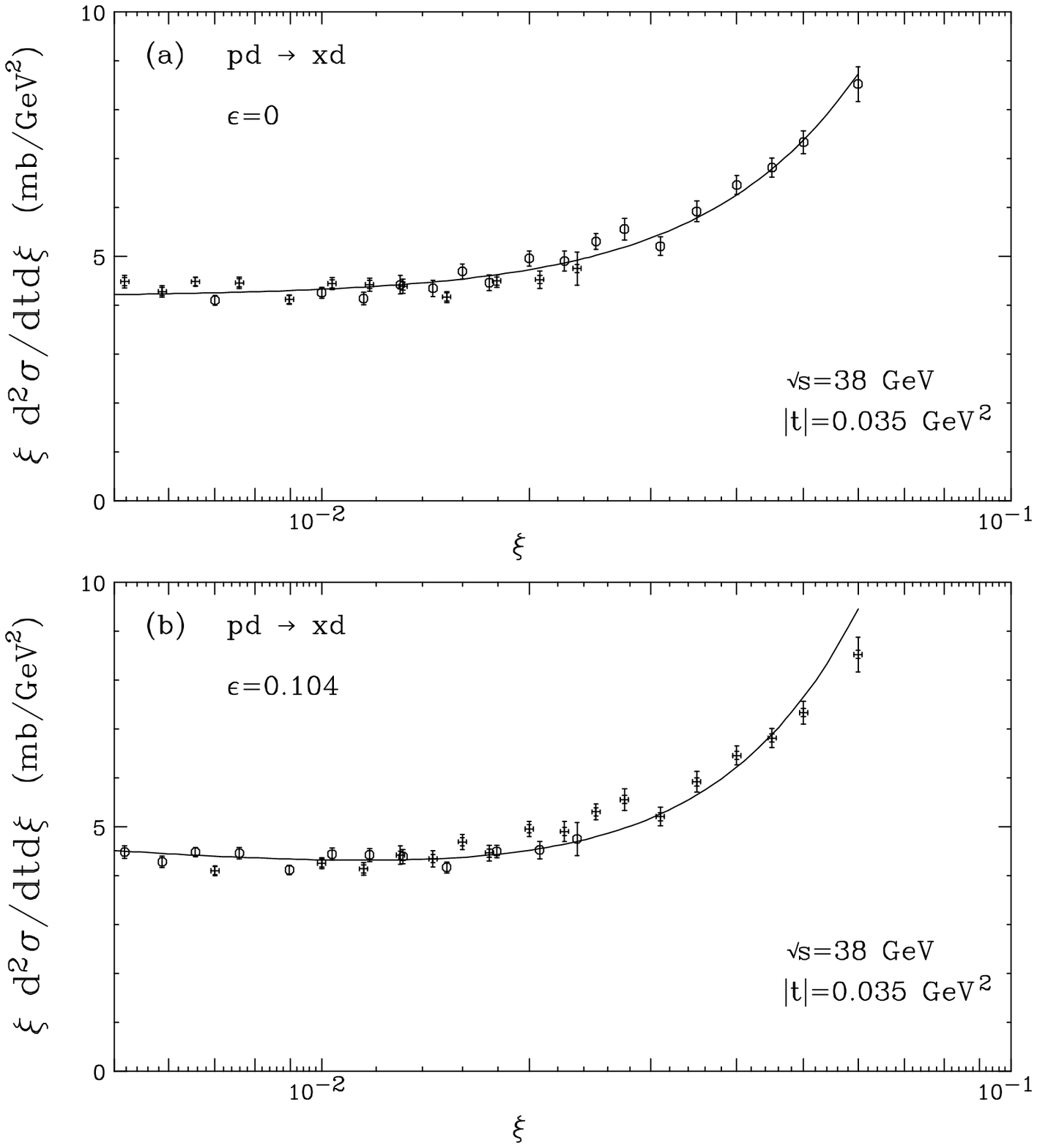}}
\vglue 0.5in
\centerline{Figures~5 (a,b): Cross sections for $p+d\rightarrow X+d$}
\clearpage
\newpage
\begin{center}
{\bf\LARGE Total single diffraction cross section for 
$p(\bar p)+p\rightarrow p(\bar p)+X$\\
(comparison with shadowing predictions)}
\end{center}
\vspace*{-0.5in}
\centerline{\psfig{figure=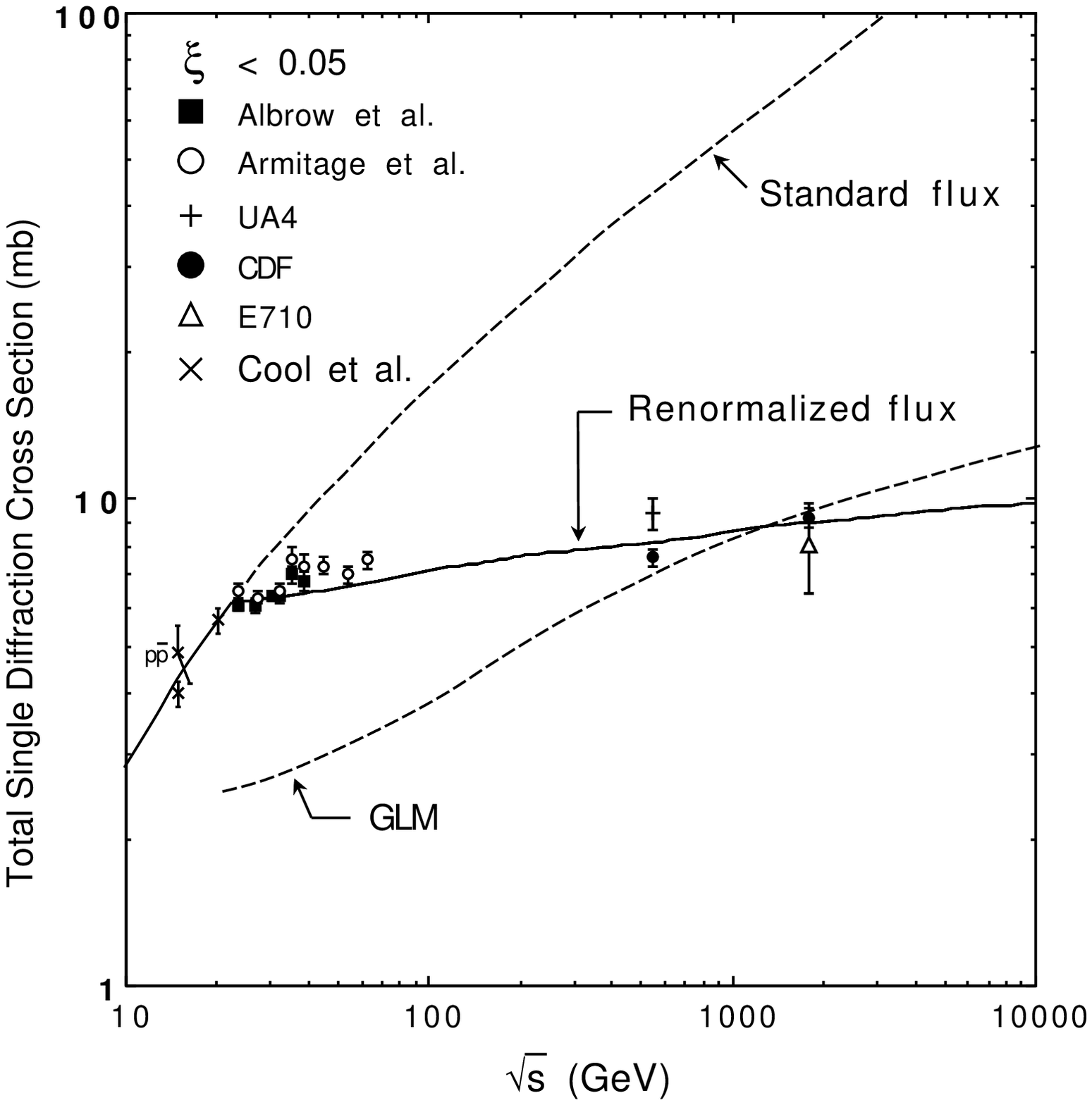,width=6in}}
\vspace*{-1.5in}
Figures~6: The total single diffraction cross section for 
$p(\bar p)+p\rightarrow p(\bar p)+X$ vs $\sqrt s$ compared with the
predictions of the renormalized pomeron flux model of Goulianos~\cite{R} 
(solid line) 
and of the model of Gotsman, Levin and Maor~\cite{GLM} 
(dashed line, labeled GLM); the latter, which includes ``screening 
corrections", 
is normalized to the 
average value of the two CDF measurements at $\sqrt s=546$ and 1800 GeV.
\clearpage
\newpage
\begin{center}
{\bf\LARGE The pomeron flux as a function of $\sqrt s$}\\
\vglue 1in
\end{center}
{\hspace*{-0.5cm}\centerline{\psfig{figure=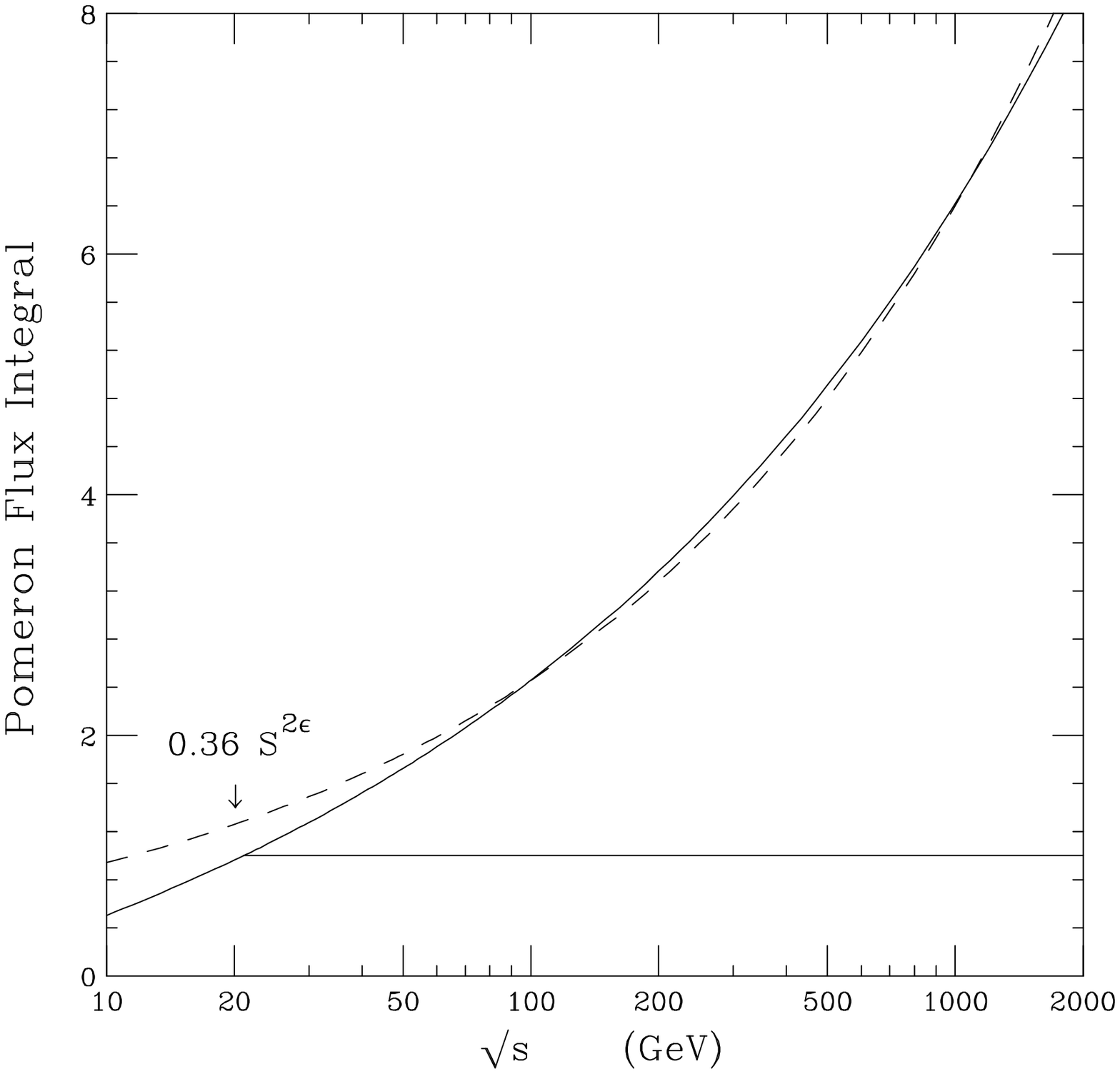,width=6in}}}

Figure~7 : The integral of the standard pomeron flux for $pp\rightarrow pX$, 
$N(s)$ of Eq.~\ref{fluxI} using $F^2(t)=e^{4.6\,t}$,  
as a function of $\sqrt s$ (solid curve) is compared with a 
dependence $\sim s^{2\epsilon}$ (dashed curve). The horizontal solid 
line at $N(s)=1$ represents the {\em saturated} renormalized flux.
The flux integral calculated using in Eq.~\ref{fluxI} the $F_1(t)$ 
form factor of Eq.~\ref{F1} can be approximated by the expression 
$0.41\,s^{2\epsilon}$.
\clearpage
\newpage
\begin{center}
{\bf\LARGE Cross sections $d^2\sigma_{sd}/d\xi dt$ at 
$t=-0.05$ GeV$^2$}\\
\vglue 1cm
\end{center}
\centerline{\psfig{figure=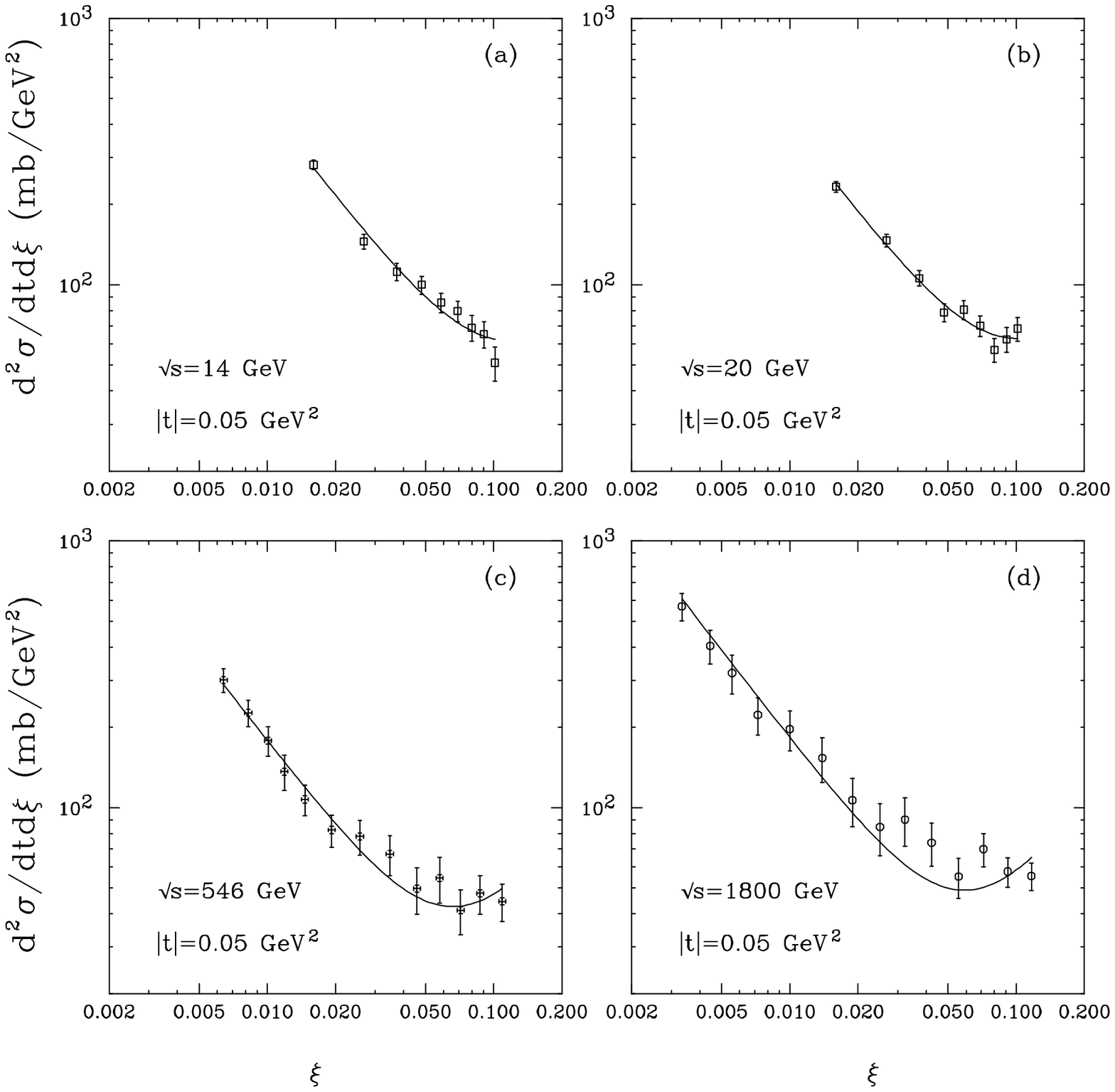,width=6in}}
\vglue 0.25in
Figure~8 : 
Cross sections $d^2\sigma_{sd}/d\xi dt$ 
for $p+p(\bar p) \rightarrow p(\bar p)+X$ at
$t=-0.05$ GeV$^2$ and $\sqrt s=14$, 20, 546 and 1800 GeV. 
The solid lines represent the best fit 
to the data at each energy using two terms, the $\pom\pom\pom$ 
and $\pi\pi\pom$ amplitudes, with their
normalizations treated as free parameters.
\clearpage
\newpage
\begin{center}
{\bf\LARGE Cross sections $\xi\cdot d^2\sigma_{sd}/d\xi dt$ at 
$t=-0.05$ GeV$^2$}\\
\vglue 1cm
\end{center}
\centerline{\psfig{figure=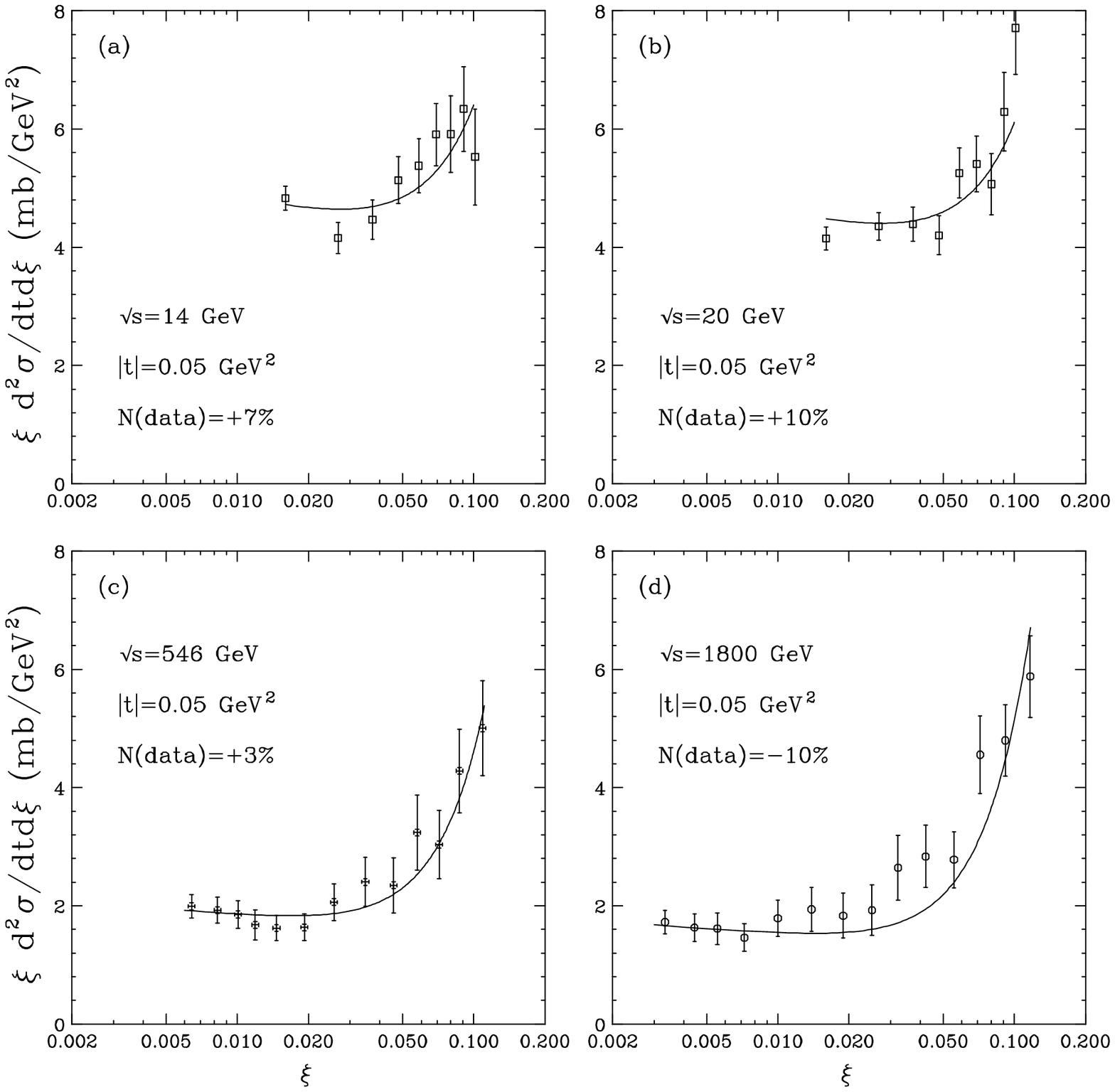,width=6in}}
Figure~9 : 
Cross sections $\xi\cdot d^2\sigma_{sd}/d\xi dt$
for $p+p(\bar p) \rightarrow p(\bar p)+X$ at
$t=-0.05$ GeV$^2$ and $\sqrt s=14$, 20, 546 and 1800 GeV
are compared with the results (solid lines)
of a simultaneous one parameter fit with a renormalized $\pom\pom\pom$
amplitude and a pion exchange contribution.
To account for systematic uncertainties,
the normalization of each data set was allowed to vary within
$\pm 10\%$ of its nominal value; the parameter ``N(data)"
represents the shift in the
data normalization for which the best fit was obtained.
\clearpage
\newpage
\begin{center}
{\bf\LARGE Cross sections $d^2\sigma_{sd}/d\xi dt$ at 
$t=-0.05$ GeV$^2$}\\
\vglue 1cm
\end{center}
\centerline{\psfig{figure=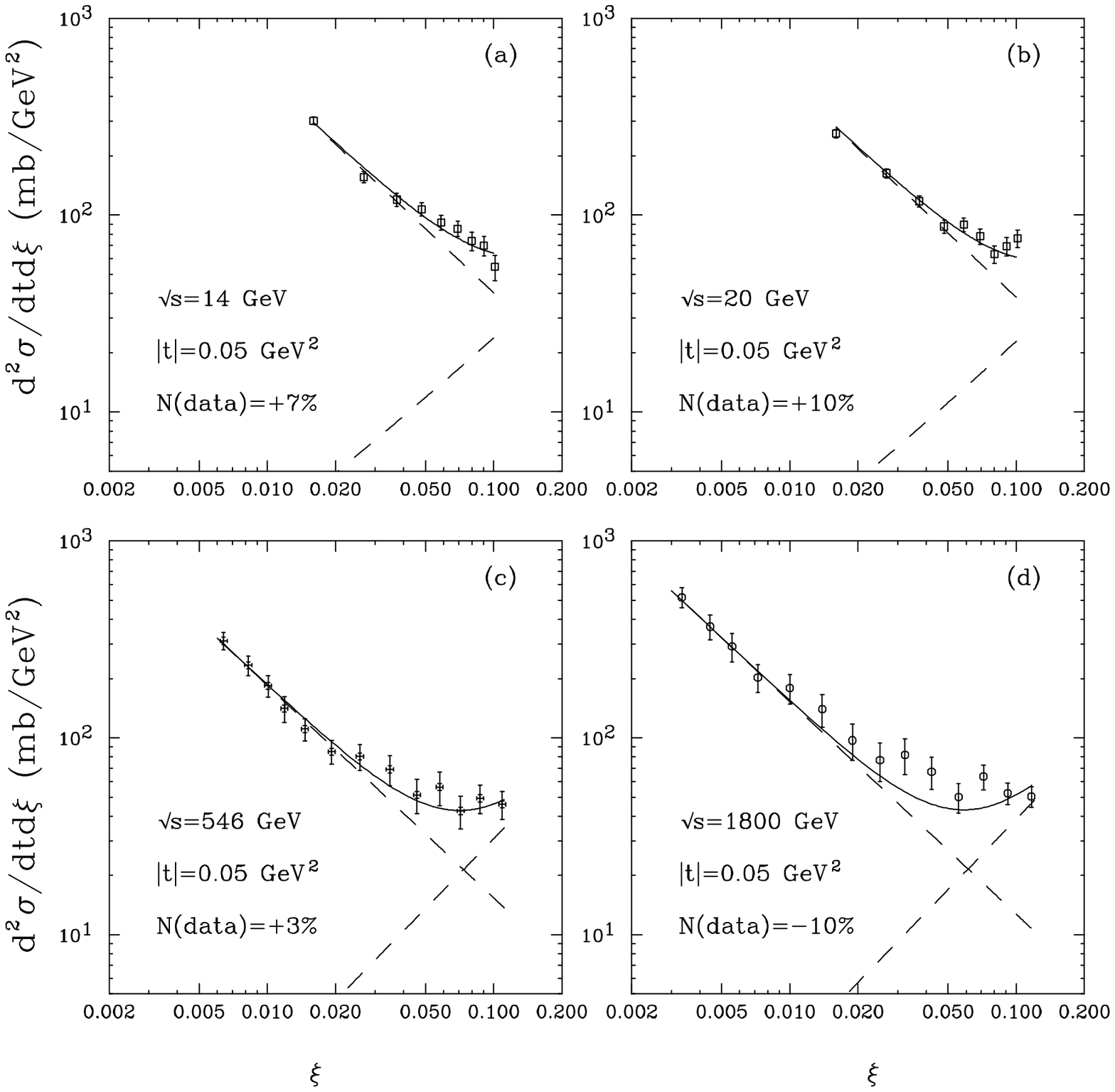,width=6in}}
\vglue 0.5cm
Figure~10 : 
Cross sections $d^2\sigma_{sd}/d\xi dt$
for $p+p(\bar p) \rightarrow p(\bar p)+X$ at
$t=-0.05$ GeV$^2$ and $\sqrt s=14$, 20, 546 and 1800 GeV
are compared with the results (solid lines)
of a simultaneous one parameter fit with
a renormalized $\pom\pom\pom$
amplitude and a pion exchange contribution.
The dashed lines represent the individual pomeron and pion contributions.
To account for systematic uncertainties,
the normalization of each data set was allowed to vary within
$\pm 10\%$ of its nominal value; the parameter ``N(data)"
represents the shift in the
data normalization for which the best fit was obtained.
\clearpage
\newpage
\begin{center}
{\bf\LARGE ISR cross sections $d^2\sigma_{sd}/d\xi dt$}\\
\vglue 1cm
\end{center}
\centerline{\psfig{figure=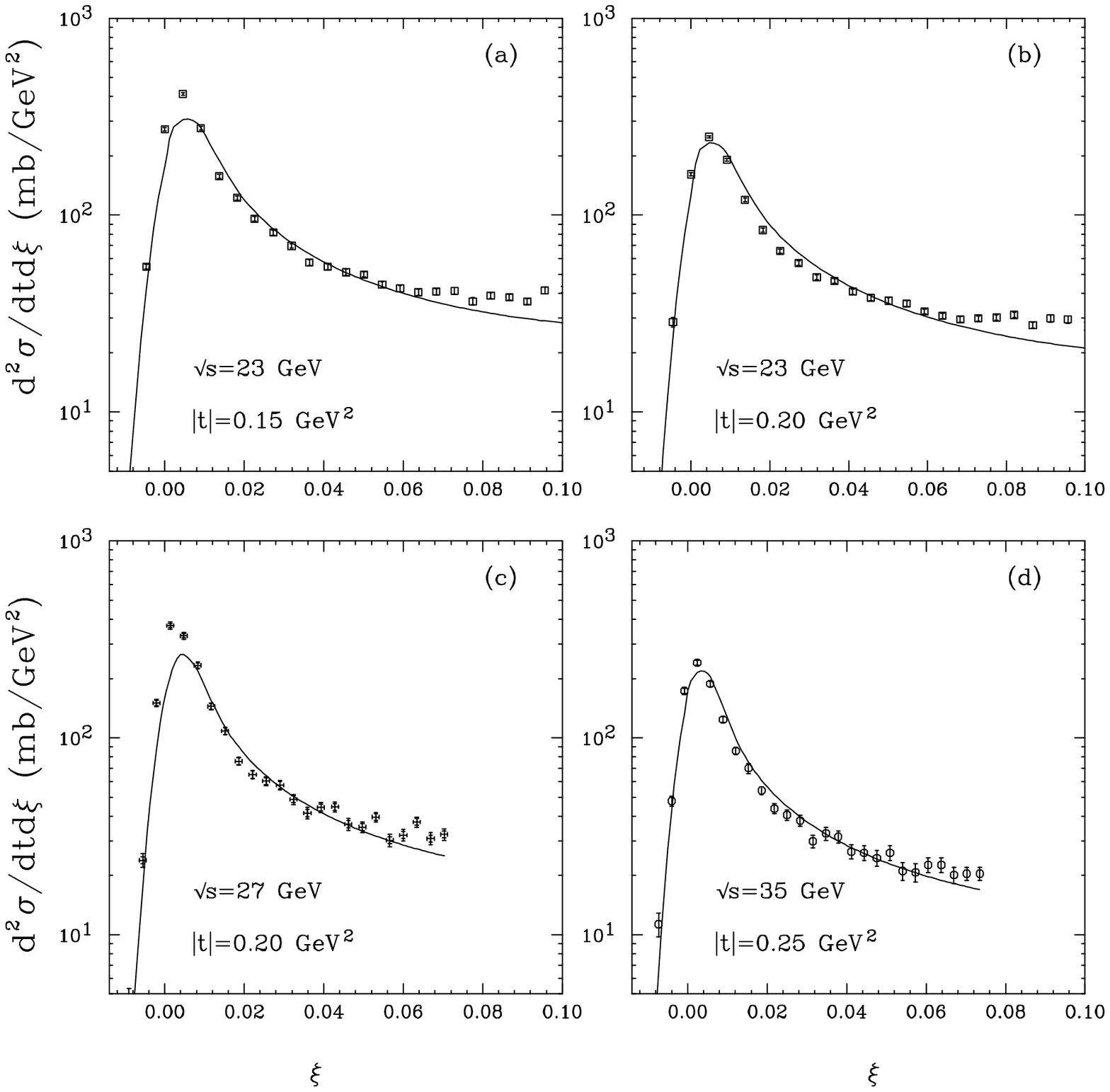,width=6in}}
\vglue 0.25in
Figure~11 : 
Cross sections $d^2\sigma_{sd}/d\xi dt$ for $pp\rightarrow pX$
measured at the ISR at various values of $\sqrt s$ and $t$.
The solid lines are fits obtained using the renormalized $\pom\pom\pom$
amplitude and the pion exchange contribution, convoluted with the
experimental $\xi$-resolution, which dominates the shape of the distributions
at small $\xi$.
The overall
normalization of the data has a systematic uncertainty of 15\%~\cite{ALBROW}.
\clearpage
\newpage
\begin{center}
{\bf\LARGE Total single diffraction cross section for
$p(\bar p)+p\rightarrow p(\bar p)+X$\\
(comparison with theoretical predictions)}
\end{center}
\vglue 0.25in
{\hspace*{-1cm}\centerline{\psfig{figure=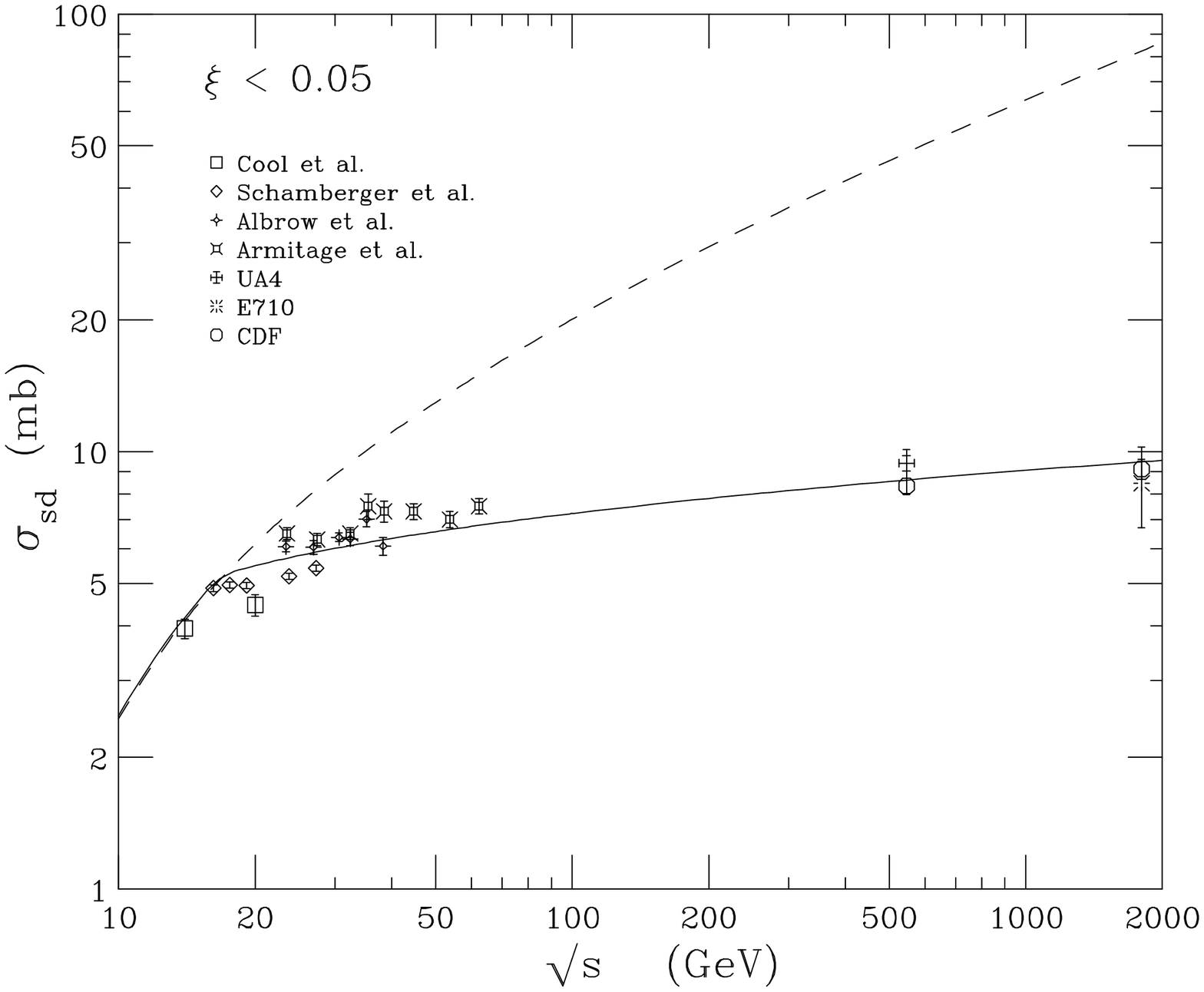,width=6in}}}
\vspace*{-3cm}

Figures~12: Total single diffraction cross sections for 
$p(\bar p)+p\rightarrow p(\bar p)+X$ versus 
$\sqrt s$ compared with triple-pomeron
predictions based (a) on pomeron pole dominance in standard Regge theory 
(dashed line) and (b) on the renormalized pomeron flux model~\cite{R} 
(solid line). The cross sections 
were corrected for effects due to extrapolations in $t$,
as discussed in the text. The errors shown are statistical; 
typical systematic uncertainties for 
each experiment are of ${\cal O}(10\%)$.
\clearpage
\newpage
\begin{center}
{\bf\LARGE Cross sections $d^2\sigma_{sd}/d\xi dt$ at $t=-0.05$ GeV$^2$}\\
\vglue 1cm
\end{center}
\centerline{\psfig{figure=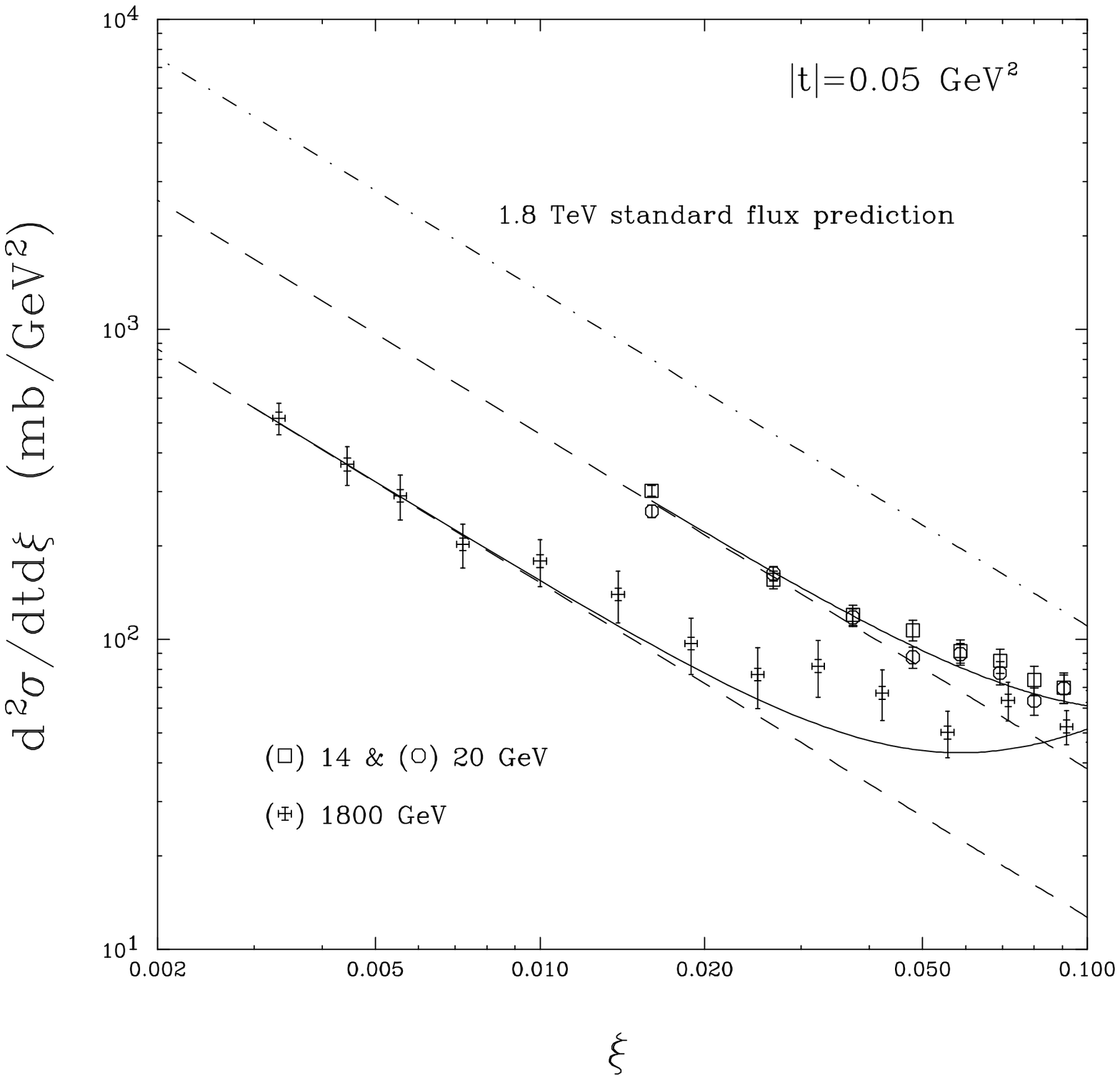,width=6in}}
Figure~13 : 
Cross sections $d^2\sigma_{sd}/d\xi dt$ 
for $p+p(\bar p) \rightarrow p(\bar p)+X$ at
$t=-0.05$ GeV$^2$ and $\sqrt s=14$, 20 and 1800 GeV. 
The solid lines are the global one-parameter fit to the data 
presented in Fig.~10, and the dashed lines represent the 
renormalized triple-pomeron contribution. The dashed-dotted line 
is the standard flux triple-pomeron contribution at $\sqrt s=1800$ GeV 
predicted from the data at $\sqrt s=14$ and 20 GeV.
\clearpage
\newpage
\begin{center}
{\bf\LARGE Cross sections $d^2\sigma_{sd}/dM^2 dt$ at $t=-0.5$ GeV$^2$}\\
\vglue 1cm
\end{center}
\centerline{\psfig{figure=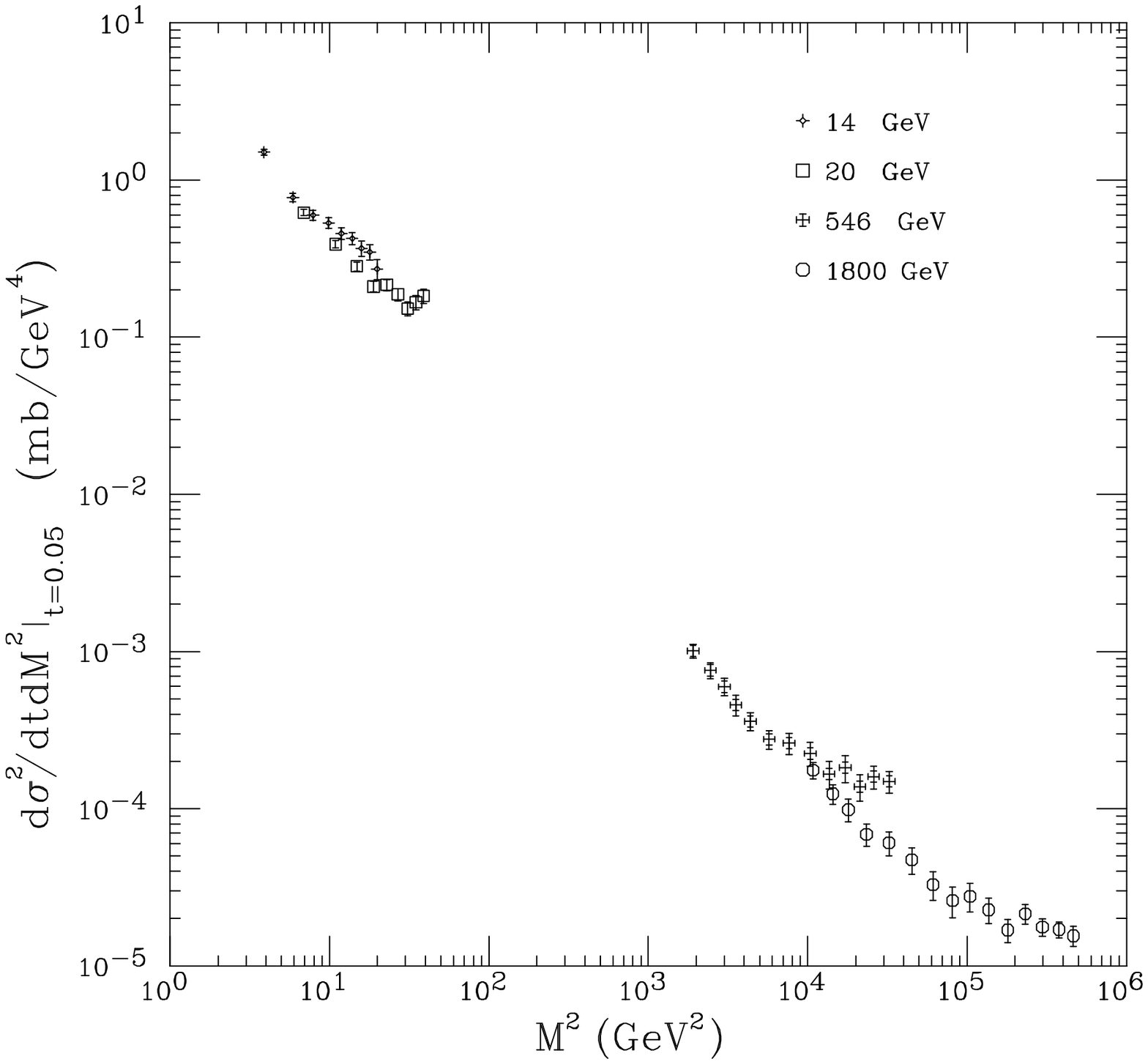,width=6in}}
Figure~14 : Cross sections $d^2\sigma_{sd}/dM^2 dt$ 
for $p+p(\bar p) \rightarrow p(\bar p)+X$ at
$t=-0.05$ GeV$^2$ and $\sqrt s=14$, 20, 546 and 1800 GeV. 
\clearpage
\newpage
\begin{center}
{\bf\LARGE Cross sections $d^2\sigma_{sd}/dM^2 dt$ at $t=-0.5$ GeV$^2$}\\
\vglue 1cm
\end{center}
\centerline{\psfig{figure=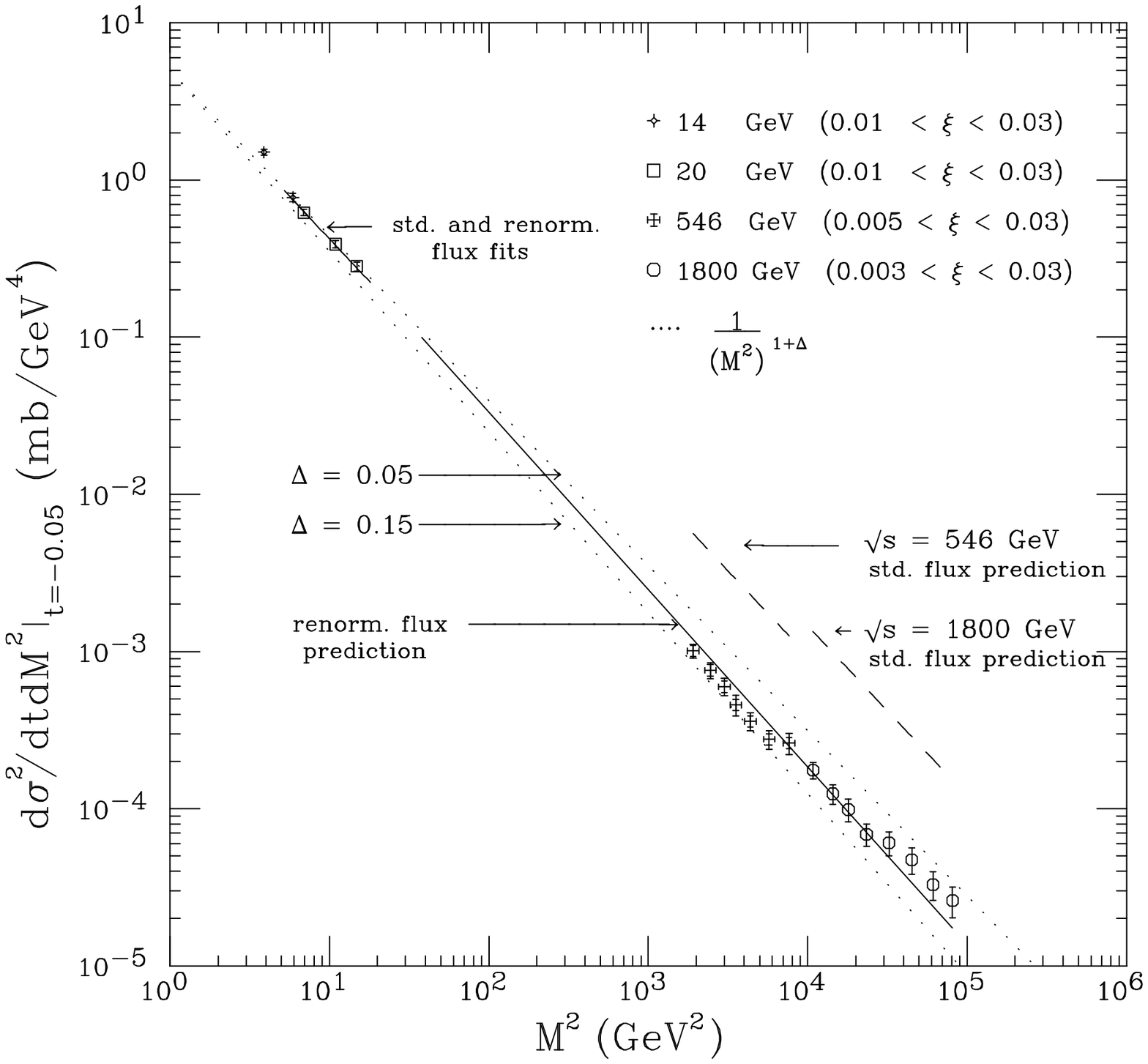,width=6in}}
Figure~15 : Cross sections $d^2\sigma_{sd}/dM^2 dt$ 
for $p+p(\bar p) \rightarrow p(\bar p)+X$ at
$t=-0.05$ GeV$^2$ and $\sqrt s=14$, 20, 546 and 1800 GeV. 
At $\sqrt s$=14 and 20 GeV,
the standard and renormalized flux fits have the same normalization; 
at the higher energies, the renormalized and standard flux predictions
are shown by the 
solid and dashed lines, respectively.
\clearpage
\newpage
\begin{center}
{\bf\LARGE Cross sections $d^2\sigma_{sd}/dM^2 dt$ at $t=0$\\
(comparison with theoretical predictions)}\\
\end{center}
\centerline{\psfig{figure=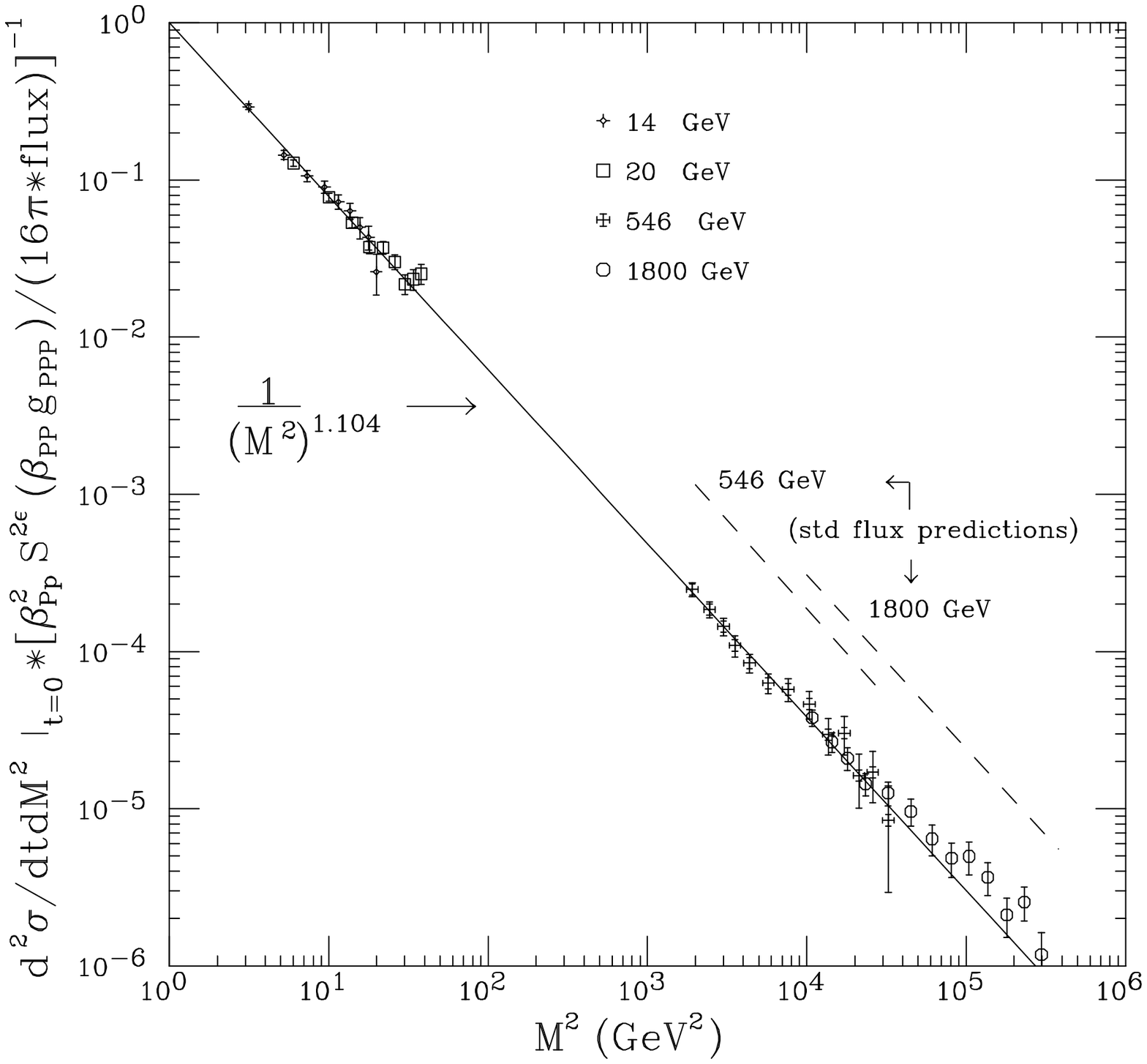,width=6in}}
Figure~16 : 
Cross sections $d^2\sigma_{sd}/dM^2 dt$
for $p+p(\bar p) \rightarrow p(\bar p)+X$ at
$t=0$ and $\sqrt s=14$, 20, 546 and 1800 GeV, multiplied by
$N(s)/[\beta^2_{\pom pp}\,s^{2\epsilon}(\beta_{\pom pp}
g_{\pom\pom\pom})/16\pi]$, are compared with the renormalized flux
prediction of $1/(M^2)^{1+\epsilon}$. The dashed curves show the
standard flux predictions.
The $t=0$ data were obtained by
extrapolation from their $t=-0.05$ GeV$^2$ values after subtracting the
pion exchange contribution.
\clearpage
\end{document}